\documentclass[12pt]{article}

\usepackage{amsgen,amsmath,amsthm,amsfonts,amstext,amsbsy,amsopn,amssymb,soul,subfigure}
\usepackage{verbatim}
\usepackage{fancyvrb}
\usepackage{graphicx}
\usepackage{amsmath}
\usepackage{booktabs}
\usepackage{dsfont}
\usepackage{natbib}
\usepackage{amssymb}
\usepackage{abstract}
\usepackage{hyperref}
\usepackage{cancel}
\usepackage{enumerate}
\usepackage{listings}
\usepackage{mathdots}
\usepackage{array}
\usepackage{phaistos}
\usepackage{textcomp}
\usepackage{algorithm}
\usepackage{float}
\usepackage{multirow}
\usepackage{algorithm}
\usepackage{algorithmicx}
\usepackage{algpseudocode}
\usepackage{setspace}
\usepackage{epstopdf}
\usepackage{pdflscape}
\usepackage{latexsym}
\usepackage{tikz}
\usepackage{setspace}
\usepackage{titlesec}
\usepackage[font=doublespacing]{caption}
\usepackage{etoolbox}
\usepackage{bigints}
\usepackage{blkarray}
\usepackage{algorithm}
\usepackage{algpseudocode}
\usepackage{enumitem}

\usepackage{xcolor}

\newtheorem{theorem}{Theorem}
\newtheorem{Condition}{Condition}
\newtheorem{Theorem}{Theorem}

\newtheorem{Lemma}{Lemma}
\newtheorem{Corollary}{Corollary}
\newtheorem{Proposition}{Proposition}
{
\theoremstyle{definition}

\newtheorem{remark}{Remark}
}

\newcommand{\be}{\begin{equation}}
\newcommand{\ee}{\end{equation}}
\newcommand{\beq}{\begin{equation}}
\newcommand{\eeq}{\end{equation}}
\newcommand{\beas}{\begin{eqnarray*}}
	\newcommand{\eeas}{\end{eqnarray*}}
\newcommand{\bea}{\begin{eqnarray}}
\newcommand{\eea}{\end{eqnarray}}
\newcommand{\bei}{\begin{itemize}}
	\newcommand{\eei}{\end{itemize}}
\newcommand{\ben}{\begin{enumerate}}
	\newcommand{\een}{\end{enumerate}}
\newcommand{\bet}{\begin{theorem}}
	\newcommand{\eet}{\end{theorem}}
\newcommand{\bel}{\begin{Lemma}}
	\newcommand{\eel}{\end{Lemma}}
\newcommand{\bep}{\begin{Proposition}}
	\newcommand{\eep}{\end{Proposition}}
\newcommand{\bed}{\begin{Definition}}
	\newcommand{\eed}{\end{Definition}}
\newcommand{\bec}{\begin{Corollary}}
	\newcommand{\eec}{\end{Corollary}}
\newcommand{\bex}{\begin{Example}}
	\newcommand{\eex}{\end{Example}}

\newcommand{\rank}{{\rm rank}}
\newcommand{\Var}{{\rm Var}}

\newcommand{\1}{{\mathbf{1}}}

\newcommand{\E}{{\mathbb{E}}}

\hypersetup{
    colorlinks=true,
    linkcolor=black,
    filecolor=magenta,      
    urlcolor=cyan,
    citecolor = blue,
    pdftitle={Overleaf Example},
    pdfpagemode=FullScreen,
    }

\begin{document}

\begin{title}
{\Large\bf Integrated Analysis for Electronic Health Records with Structured and Sporadic Missingness}

\author{Jianbin Tan\thanks{Department of Biostatistics \& Bioinformatics, Duke University, NC, USA.}~$^1$, ~ Yan Zhang\thanks{Department of Biostatistics \& Bioinformatics, Duke University, NC, USA.}~$^1$, ~ Chuan Hong\thanks{Department of Biostatistics \& Bioinformatics, Duke University, NC, USA}, ~ T. Tony Cai\thanks{Department of Statistics \& Data Science, University of Pennsylvania, PA, USA},\\ ~ Tianxi Cai\thanks{Department of Biostatistics and Department of Biomedical Informatics, Harvard University.}~$^2$, ~ and ~ Anru R. Zhang\thanks{Department of Biostatistics \& Bioinformatics and Department of Computer Science, Duke University, NC, USA}~$^2$}
\date{}
\end{title}

\newpage
\clearpage

\begin{sloppypar}
\maketitle

\footnotetext[1]{These authors contributed equally.}
\footnotetext[2]{Joint corresponding authors.}

\begin{abstract}
Objectives: We propose a novel imputation method tailored for Electronic Health Records (EHRs) with structured and sporadic missingness. Such missingness frequently arises in the integration of heterogeneous EHR datasets for downstream clinical applications. By addressing these gaps, our method provides a practical solution for integrated analysis, enhancing data utility and advancing the understanding of population health.

Materials and Methods: We begin by demonstrating structured and sporadic missing mechanisms in the integrated analysis of EHR data. Following this, we introduce a novel imputation framework, \textsc{Macomss}, specifically designed to handle structurally and heterogeneously occurring missing data. We establish theoretical guarantees for \textsc{Macomss}, ensuring its robustness in preserving the integrity and reliability of integrated analyses. To assess its empirical performance, we conduct extensive simulation studies that replicate the complex missingness patterns observed in real-world EHR systems, complemented by validation using EHR datasets from the Duke University Health System (DUHS).

Results: Simulation studies show that our approach consistently outperforms existing imputation methods. Using datasets from three hospitals within DUHS, \textsc{Macomss} achieves the lowest imputation errors for missing data in most cases and provides superior or comparable downstream prediction performance compared to benchmark methods.

Discussion: The proposed method effectively addresses critical missingness patterns that arise in the integrated analysis of EHR datasets, enhancing the robustness and generalizability of clinical predictions.

Conclusions: We provide a theoretically guaranteed and practically meaningful method for imputing structured and sporadic missing data, enabling accurate and reliable integrated analysis across multiple EHR datasets. The proposed approach holds significant potential for advancing research in population health.
\end{abstract}

\noindent%

\section{Background and Significance}
\label{sec.intro}

Electronic Health Records (EHRs) have become a cornerstone of modern healthcare, offering rich, multidimensional data that support clinical decision-making, advance scientific research, and guide health policy development \citep{evans2016electronic,essen2018patient,beaulieu2018characterizing,tayefi2021challenges,ahuja2022mixehr,psychogyios2023missing,tian2024reliable,tan2024functional,li2024mixehr}.  
With the widespread adoption of EHR systems, the volume and diversity of collected data have grown significantly \citep{hemingway2018big,johnson2020mimic}, creating great potential for integrated analyses to deepen our understanding of population health.  
Yet, due to the intricate nature of healthcare delivery and data collection practices, we inevitably encounter challenges related to missing data \citep{madden2016missing,beaulieu2018characterizing,hemingway2018big,haneuse2021assessing,tan2022transmission,psychogyios2023missing,luo2025functional}.  
These gaps not only hinder the ability to derive accurate insights but also limit data utility, opening new opportunities to develop advanced methods for integrated analysis of EHR data.

When integrating datasets from multiple sources, structured and sporadic missingness are two common missing mechanisms that are often encountered in EHR analysis.  
Structured missingness typically results from the fact that different data sources are collected from distinct patient populations.  
For example, due to resource constraints or disease-specific clinical workflows, not all tests or procedures are performed and recorded uniformly across data sources \citep{madden2016missing, bower2017addressing, haneuse2021assessing}.  
This often leads to systematic gaps and misalignment when integrating data into a unified structure.  
In addition, disparities in data accessibility across organizations—driven by factors such as privacy regulations, institutional policies, or differences in technological infrastructure \citep{beard2012challenges, keshta2021security, tertulino2024privacy}—may further contribute to such structured gaps in the integrated analysis.

Sporadic missingness, by contrast, often arises from incomplete documentation, technical malfunctions, or missing responses to surveys \citep{wells2013strategies,bower2017addressing,haneuse2021assessing}. This type of missingness may occur randomly, with mechanisms that vary not only across datasets but also within a single dataset.  
The coexistence of structured and sporadic missingness poses significant challenges for EHR data analysis; these difficulties may be further compounded by the heterogeneous nature of data across healthcare systems.  
These challenges highlight the need for advanced imputation methods capable of accommodating diverse missingness patterns in integrated analyses.

Recently, various imputation methods have been developed to address missing data in EHR datasets, ranging from traditional statistical techniques to advanced machine learning approaches. Classical methods, such as mean imputation \citep{little2019statistical} or regression imputation \citep{JSSv045i03,pmm_cite}, are straightforward but often struggle to handle the complexity of missingness in EHR data.
{Machine learning-based techniques, such as random forests, K-NN, principal component analysis, support vector machines, matrix factorization, and deep learning imputation methods \citep{wang2015rubik, madden2016missing, beaulieu2017missing, beaulieu2018characterizing, nazabal2020handling, you2020handling, yoon2020vime, li2021imputation, aidos2021neighborhood,pathak2022imputing, psychogyios2023missing,tan2024graphical}, have shown promise in imputing missing data by leveraging the richness of available information.}
However, these approaches usually require that data are missing at random or uniformly distributed across the dataset.
This assumption may not be reasonable and is often violated in EHR data \citep{wells2013strategies,beaulieu2018characterizing,haneuse2021assessing,getzen2023mining}, making such methods less effective in addressing the distinct mechanisms of structured and sporadic missingness.

\section{Objective}
\begin{figure}[h]
    \begin{center}
        \includegraphics[scale = 0.4]{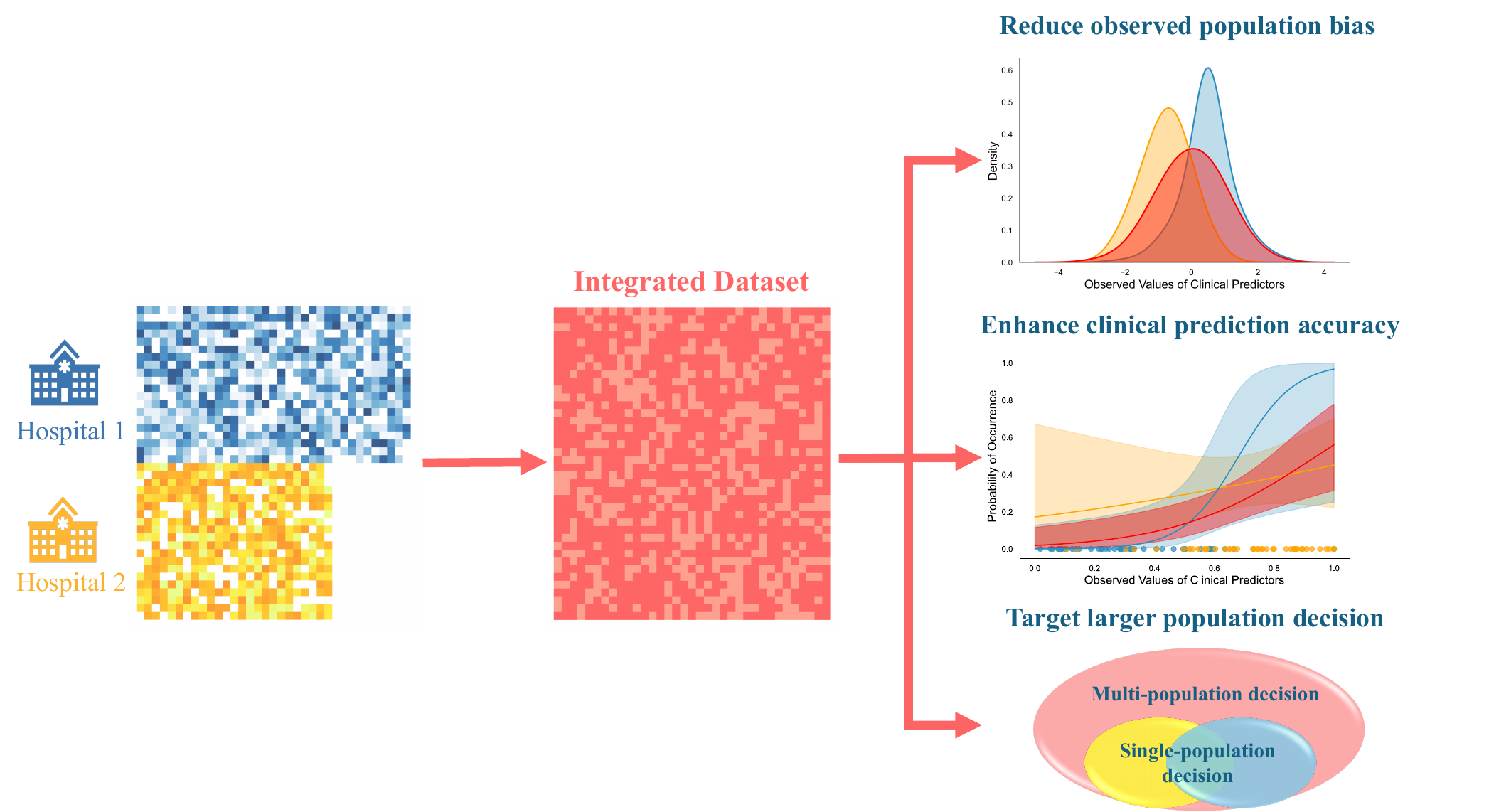}
    \caption{An illustration of the objective of this article.
    }
    \label{Objective}
    \end{center}
\end{figure}

In this article, we propose a novel data imputation method tailored for datasets with structured and sporadic missingness, paving the way for integrating multiple EHR datasets for their downstream tasks in clinical analysis. 
Our goal is to develop a robust and theoretically grounded approach that effectively mitigates both systematic and randomly occurring gaps in integrated analysis.
By addressing these complexities, we improve the completeness and accuracy of EHR datasets, supporting more precise clinical insights and population health research. A flowchart illustrating our objective is presented in Figure~\ref{Objective}.

\section{Materials and Methods}
\label{sec.procedure}

In this section, we present a detailed procedure for data imputation of structured and sporadic missingness. We begin with basic notations and definitions that will be used throughout the paper.
For any two real numbers $a$ and $b$, we note $a\wedge b$ and $a\vee b$ as the minimum and the maximum of $a$ and $b$, respectively. 
Denote $\|\cdot\|_F$ and $\|\cdot\|$ as the Frobenius and spectral norm of a matrix; their definitions are given in Part~\ref{sec.theory} in Supplementary Materials.

\subsection{Matrix Completion with Structured and Sporadic Missing Data}\label{sec:mc-structured-sporadic-missing}

We particularly focus on the following setting in this paper. 
For a high-dimensional low-rank matrix \( A \in \mathbb{R}^{p_1 \times p_2} \), we observe its \( m_1 \) rows and \( m_2 \) columns with noise and possible missing values. Specifically, let \( Y = A + Z \in \mathbb{R}^{p_1 \times p_2} \) represent the potential noisy observations without missingness, and \( Z \in \mathbb{R}^{p_1 \times p_2} \) denote the noise. To analyze the missingness, we introduce \( M \in \mathbb{R}^{p_1 \times p_2} \) as the indicator for observable/missing entries:
\[
Y_{ij} \text{ is observed if } M_{ij} = 1; \quad Y_{ij} \text{ is missing if } M_{ij} = 0.
\]

Without loss of generality, we permute the structurally missing block to the bottom right corner, and $Y$ and $M$ can be written in the following block form:
\begin{equation}\label{eq:Y_block}
Y = \begin{blockarray}{ccc}
m_2 & p_2 - m_2 & \\
\begin{block}{[cc]c}
Y_{(11)} & Y_{(12)} & m_1 \\
Y_{(21)} & \color{gray}{Y_{(22)}} & p_1-m_1\\
\end{block}
\end{blockarray}, \quad M = \begin{blockarray}{ccc}
m_2 & p_2 - m_2 & \\
\begin{block}{[cc]c}
M_{(11)} & M_{(12)} & m_1 \\
M_{(21)} & M_{(22)} & p_1-m_1\\
\end{block}
\end{blockarray}.
\end{equation}
The block \( Y_{(22)} \) is unobserved, so all entries of \( M_{(22)} \) are zero. Moreover, entrywise missingness might exist in \( Y_{(11)} \), \( Y_{(12)} \), and \( Y_{(21)} \), meaning that a subset of the entries in \( M_{(11)} \), \( M_{(12)} \), and \( M_{(21)} \) are zero. To model the sporadic missingness, we assume that each entry in the observable rows and columns is missing independently according to a probability matrix \( \Theta \).  In other words, \( M_{ij} \sim \text{Bernoulli}(\Theta_{ij}) \) for $i, j$ corresponding to the entries of $M_{(11)}, M_{(12)}, M_{(21)}$.

Here, the rows of \( Y \) usually represent patients from multiple EHR systems, while the columns indicate the observed clinical features of the patients. The unobserved block \( Y_{(22)} \) may arise when the observed entries in each row are misaligned. 
This misalignment often occurs in EHR integrated analysis, where the observed clinical features may differ across different sources of datasets \citep{madden2016missing,bower2017addressing,haneuse2021assessing}.

The block-structured missingness in \eqref{eq:Y_block} introduces significant challenges in estimating the latent probability matrix \( \Theta \).
{To address this difficulty, we focus on the case of a rank-one probability matrix, i.e., \( \text{rank}(\Theta) = 1 \). The rank-one missingness model has been widely studied in the matrix completion literature (see, e.g., \cite{keshavan2010matrix, chatterjee2015matrix, cho2015asymptotic}), which allows the sporadic missingness in \( Y \) to occur heterogeneously across patients, clinical features, and data sources.
This assumption offers a parsimonious structure and facilitates both algorithmic design and theoretical analysis, but it does not need to be strictly satisfied in practice for the methods to perform well (see Section~\ref{sec.real_data} for a demonstration).}

{Our goal is to recover \( A \) in an unsupervised manner from the available observations \( Y_{ij} \) for those \( (i,j) \) pairs where \( M_{ij} = 1 \)}, considering both the structured and sporadic missingness mechanisms in the data.
{Clearly, such a task is impossible in general without further assumptions. Motivated by applications in matrix completion, we assume the original matrix \( A \) is low-rank (or nearly low-rank), which enables our method to impute missing values by leveraging low-dimensional shared structure.}
We postpone more detailed discussions on technical conditions to Part \ref{sec.theory} in Supplementary Materials.

\subsection{Algorithm}\label{sec:algo}

The proposed algorithm for matrix completion with structured and sporadic missing values consists of five steps and two key components: a spectral method for imputing sporadic missingness and denoising (Steps 1 and 2), and a structured matrix completion method for imputing structured missingness (Steps 3, 4, and 5). 
The central ideas behind Steps~1 and~2 are mean matching for aligning the observations and parameters, and singular value decomposition for extracting low-rank structures from missing entries and noise. 
Steps~3,~4, and~5 rely on a simple yet powerful algebraic fact—the Schur complement. 
Below, we outline the five steps for estimating the noiseless matrix $A$ in detail:

\begin{enumerate}[leftmargin=*]

\item[Step 1.] (Estimation of Missingness Parameter $\Theta$)
Without loss of generality, we permute the structurally missing block to the bottom right corner and assume that $Y$ and $M$ are written in the block structure form \eqref{eq:Y_block}.
In this step, we estimate the missingness parameters, denoted by $\Theta_{ij}$, for all observable rows and columns. Given the assumption that the parameter matrix $\Theta$ is rank-one, Lemma \ref{lm:rank-one-matrix-expression} in Supplementary Materials indicates that each entry of $\Theta$ can be represented as the product of corresponding row and column sums divided by the total sum of the entries in $\Theta$ (i.e., $\Theta_{ij}= (\sum_{i'=1}^{p_1} \Theta_{i'j})(\sum_{j'=1}^{p_2} \Theta_{ij'})/(\sum_{i'j'}\Theta_{i'j'})$). Notice that the expectation of $M_{ij}$ is $\Theta_{ij}$ (i.e., \( \mathbb{E} M_{ij} = \Theta_{ij} \)), we estimate \( \Theta_{ij} \) by computing the corresponding products of the row and column sums of \( M \), normalized by the overall sum of the entries in $M$.

\item[Step 2.] (Imputation of Missing Values)
We begin by filling all missing entries of the original data matrix $Y$ with zeros. Subsequently, we normalize each observed entry by dividing it by the corresponding estimated missingness parameter from Step 1 (i.e., $\tilde{Y}_{ij}:=Y_{ij}/\hat\Theta_{ij}$). This procedure yields a normalized matrix $\tilde{Y}$, which removes biases introduced by missing observations. We then partition the normalized data matrix into four sub-blocks following \eqref{eq:Y_block}, denoted by $\tilde{Y}_{(11)}, \tilde{Y}_{(12)}$, $\tilde{Y}_{(21)}$, and $\tilde{Y}_{(22)}$. Notably, all entries of $\tilde{Y}_{(22)}$ are zeros. 

\item[Step 3.] (Rotation of $\tilde{Y}_{(\bullet 1)}, \tilde{Y}_{(1\bullet)}$)
Notice that $\tilde{Y}_{(11)}$, $\tilde{Y}_{(12)}$, and $\tilde{Y}_{(21)}$ are noise observations of the low-rank matrix $A_{(11)}$, $A_{(12)}$, and $A_{(21)}$, we propose to apply the spectral method on $\tilde{Y}$ to extract the underlying low-rank structure.
Specifically, we calculate the singular value decomposition for 
$$Y^{\rm (col)} = \begin{bmatrix}
\frac{(p_1-2m_1)\wedge 0}{p_1}\cdot \tilde{Y}_{(11)} \\
\tilde{Y}_{(21)}
\end{bmatrix}, \quad \text{and}\quad Y^{\rm (row)} = \left[\frac{(p_2-2m_2)\wedge 0}{p_2}\cdot \tilde{Y}_{(11)}, ~ \tilde{Y}_{(12)}\right].$$ 
The resulting singular vectors provide rotation matrices for the observable blocks $\tilde{Y}_{(11)}, \tilde{Y}_{(12)}$, and $\tilde{Y}_{(21)}$, obtaining the rotated matrices $B_{(11)}, B_{(12)}, B_{(21)}$. After the rotations, the significant components hidden in $\tilde{Y}$ are moved to the front ranking rows and columns in $B_{(11)}, B_{(12)}$, and $B_{(21)}$.

\item[Step 4.] (Trimming and Rank Determination)
In this step we aim at trimming $B_{(11)}, B_{(12)}$, and $B_{(21)}$ to low-rank blocks, where the specific rank $\hat{r}$ needs to be estimated.
To do this, we iteratively estimate an appropriate rank for truncation, starting from an initial upper bound and decreasing sequentially. The optimal rank choice satisfies certain stability conditions involving bounds based on the dimensions of observable and missing data, as discussed in Part~\ref{sec.theory} in Supplementary Materials. 

\item[Step 5.] (Assembling and Imputation)
In the final step, we treat $A$ as a rank-$\hat{r}$ matrix and reconstruct $A$ by combining the trimmed blocks $B_{(11)}, B_{(12)}, B_{(21)}$ using rotations obtained in Step 3. 

\end{enumerate}

By combining the steps described above, our procedure can be implemented as outlined in Algorithm \ref{al:procedure} in Supplementary Materials, referred to as \underline{Ma}trix \underline{co}mpletion with \underline{M}issing \underline{S}tructurally and \underline{S}poradically (\textsc{Macomss}). 
Unlike conventional machine-learning-based imputation methods \citep{wang2015rubik,madden2016missing,beaulieu2017missing,beaulieu2018characterizing,li2021imputation,pathak2022imputing,psychogyios2023missing}, \textsc{Macomss} is tuning-free and straightforward to implement. In addition, the proposed approach effectively accounts for structured and sporadic missing mechanisms during imputation and incorporates a denoising process for the observed data. In contrast, the existing imputation methods often overlook such complex missing mechanisms and incorporate noise from the observations into imputation processes. This can lead to inaccuracies when the data exhibits structured and sporadic missingness and are observed with relatively high noise.

{In practice, we may encounter both continuous and categorical variables simultaneously in matrix completion tasks. For this setting, \textsc{Macomss} is applicable to both types of variables based on a sub-Gaussian framework (see Part~\ref{sec.theory} of the Supplementary Materials), which encompasses a broad class of continuous or discrete distributions such as Gaussian, uniform, Bernoulli, Binomial, Hypergeometric, and bounded discrete uniform distributions \citep{vershynin2018high}. Certain unbounded distributions, such as Poisson, may also exhibit sub-Gaussian behavior under appropriate conditions or after a log transformation. Due to this, we always apply a logarithmic transformation to count data before imputation.}

Under the sub-Gaussian framework, we analyze the statistical lower and upper bounds of \textsc{Macomss} for estimating the latent matrix $A$ from the observed data \( Y_{ij} \)s. These results not only highlight the theoretical optimality of \textsc{Macomss} in preserving the data integrity.

\subsection{Validation}
\label{sec.simulation}

\subsubsection{Simulations for Matrix Recovery}\label{sec:recorery}

We first consider the recovery performance of \textsc{Macomss} under various values of $m_1$ and $m_2$, which represent the number of observable rows and columns, respectively. 
Particularly, we fix $p_1 = p_2 = 300$, $r = 3$, and generate $A = U V^\top$, where $U \in \mathbb{R}^{p_1\times r}$ and $V \in \mathbb{R}^{p_2\times r}$ are uniformly random orthogonal matrices. We then uniformly randomly select $m_1$ rows and $m_2$ columns for observation with missing values under the following settings:

\begin{enumerate}
    \item We vary $m_1$ and $m_2$ among $\{10, 20, \ldots, 100\}$. 
The missing parameter $\Theta$ is constructed as a rank-1 matrix: $\Theta = \alpha \beta^\top$, where $\alpha \in \mathbb{R}^{p_1}$, $\alpha_i \sim 1 - 0.05 \cdot \operatorname{Unif}[0, 1]$, and $\beta \in \mathbb{R}^{p_2}$, $\beta_j \sim 1 - 0.05 \cdot \operatorname{Unif}[0, 1]$. 
Finally, we corrupt each observation with i.i.d. Gaussian noise as in $Y = A + Z$, and introduce sporadic missingness to $Y$ based on the probability matrix $M$. Here, the variance of the entries in $Z$ is given as $\sigma = 0.3 \cdot {\|A\|_F}/ {\sqrt{p_1p_2}}$, and $M$ is sampled from Bernoulli distributions with the missing probability being the entries in $\Theta$.
Such a setting mimics the real data situation in EHR datasets, where a small portion of sporadic missingness and observational noise typically exist. 
\item We consider another setting for investigating the influence of the noise $\sigma$ and the missingness $\Theta$ on the recovery performance. Specifically, we fix $m_1 = m_2 = 50$, and $A$ to be generated in the same way as the previous setting. We let $\sigma$ vary from $0.2 \cdot {\|A\|_F}/ {\sqrt{p_1p_2}}$ to $2 \cdot {\|A\|_F}/ {\sqrt{p_1p_2}}$ and set $\Theta = \alpha \cdot \beta^\top$, where $\alpha \in \mathbb{R}^{p_1}, \beta \in \mathbb{R}^{p_2}$, and $\alpha_i, \beta_j \overset{\text{iid}}{\sim} \mathrm{Unif}[1-\eta, 1]$, with $\eta$ varying from 0 to 0.25. A larger $\eta$ indicates a higher missing probability.
\end{enumerate}

All experiments in the above settings are repeated for 1,000 times.
Based on the observable entries from $(Y, M)$, we apply Algorithm \ref{al:procedure} in Supplementary Materials to obtain estimates $\hat{A}$. 
We utilize the average Frobenius and spectral norm loss of recovery to evaluate the accuracy of $\hat{A}$ for estimating $A$.

\subsubsection{Simulations for Downstream Tasks}\label{sec: DST}

Next, we evaluate the performance of \textsc{Macomss} for downstream tasks after matrix completion. Specifically, we focus on logistic regressions on the dataset $(X, Z)$, where $X \in \mathbb{R}^{n \times p}$ contains the observations of $n$ samples with $p$ features as predictors, and $Z \in \mathbb{R}^n$ represents the binary responses. Here, $n$ represents the number of patients, $p$ denotes the number of clinical features for patients, $Z$ contains binary outcomes of patients, and $X$ is the predictor matrix, including observed values of clinical features from patients.
To mimic the setting in real-world EHR datasets, we suppose that the matrix $X$ presents with both structured and sporadic missing entries and is observed with noise. Our objective is to impute the missing values of \( X \) and denoise \( X \), and then use the recovered matrix to predict the response \( Z \).

For data generation, we first sample the matrix $X$ and then generate the response $Z$ using a logistic model; see Section~\ref{sim_add} in Supplementary Materials for details.
To introduce sporadic missingness, we generate the observed values of \( X \) according to a rank-1 missing probability matrix \( \Theta \), similar to those in Section~\ref{sec:recorery}. Specifically, \( \Theta \) is constructed using \( \Theta = \alpha \beta^\top \),
where \( \alpha \in \mathbb{R}^{n} \) and \( \beta \in \mathbb{R}^{p} \). Each element of \( \alpha \) and \( \beta \) is independently drawn from \( 1 - 0.1 \cdot \text{Unif}[0, 1] \).
Given the matrix \( \Theta \), we then generate the label \( M_{ij} \) by \( M_{ij} \sim \text{Bernoulli}(\Theta_{ij}) \), where \( \Theta_{ij} \) is the \( (i,j) \)-th element in \( \Theta \). If \( M_{ij} = 1 \), we then observe the \( (i,j) \)-th element of \( X \).

Furthermore, we introduce a structured missing mechanism in the predictor matrix \( X \) under the two scenarios:
\begin{itemize}
    \item In scenario 1, we evaluate different row numbers, \( n \), and assume that the intersection of the final {60\%} of rows and the last 45 columns is completely missing, while the rest is observable with sporadic missingness. 
\item In scenario 2, we instead evaluate different column numbers, \( p \), and suppose that the intersection of the final 45 rows and the last {60\%} of columns is entirely missing, with the remaining entries observable with sporadic missingness.
\end{itemize}
These two scenarios correspond to cases where the missing block enlarges with the number of rows (patients) or columns (clinical features).

Finally, for each observed entry of \( X \), we add Gaussian noise \( \operatorname{Gau}(0, \sigma) \), 
where \( \sigma \) is set as \( \text{SNR} \times \frac{\|X\|_F}{\sqrt{np}} \), with SNR denoting the signal-to-noise ratio. 
The contaminated and incomplete observations of \( X \), together with the responses \( Z \) generated by the noiseless predictor matrix \( X \), 
are then used for estimation.

{We compare \textsc{Macomss} with several existing data imputation methods, abbreviated as PMM, BLR, RS, CART, K-NN, VAE, and VAA, where the first five methods are implemented using the \texttt{R} package \texttt{MICE} \citep{JSSv045i03}.}
Specifically, Predictive Mean Matching \citep[PMM;][]{pmm_cite} is a regression-based method that provides imputed values by predicting them through regression. Bayesian Linear Regression \citep[BLR;][]{JSSv045i03} is another regression-based method that incorporates parameter uncertainty using prior distributions. Random Sampling (RS) imputes missing values by randomly selecting from observed values within the same feature. Classification and Regression Trees \citep[CART;][]{cart_cite} use a machine learning technique that partitions the data into subsets and imputes missing values by learning from these partitions. K-Nearest Neighbors \citep[K-NN;][]{KNN_cite} is another machine learning approach that imputes data using the similarity between data points. In addition,
{Variational Autoencoders \citep[VAE;][]{nazabal2020handling} are deep generative models that learn a latent representation of the data and reconstruct missing values by sampling from this latent space. 
Variational neighborhood‑aware autoencoders \citep[VAA;][]{aidos2021neighborhood} extend VAE by coupling a latent generative model with a K‑NN, locality-driven imputation step.
}

We apply the above methods to impute the missing data in $X$ based on its observed entries. In addition, we employ \textsc{Macomss} for imputation and denoising of $X$.
Following this, we implement a logistic regression between $Z$ and the imputed matrix of $X$, utilizing elastic net regularization \citep{zou2005regularization} via the \textsf{R} package \textbf{glmnet} \citep{friedman2010regularization}. 

To evaluate the matrix recovery performance of each method, we calculate the normalized mean squared errors (NMSE) for the missing entries of $X$: 
\[
\text{NMSE}_X = \frac{\|\hat{X}_{\text{miss}}-X_{\text{miss}}\|^2}{\|X_{\text{miss}}\|^2}.
\]
Here, $X_{\text{miss}}$ and $\hat{X}_{\text{miss}}$ are vectors of the true and estimated values for the missing entries. In addition, we calculate the area under the ROC curve \citep{huang2005using}, denoted as $\text{AUC}$, to evaluate predictive performances of classification, where the classification function is estimated from the imputed predictor matrix and the responses $Z$ using logistic regressions.
For each setting, we replicate 100 simulations to calculate the NMSE and AUC. In calculating $\text{AUC}$, we also include a complete-data benchmark method where the classification function is estimated from the logistic regression of \( Z \) onto the true predictor matrix \( X \).

\subsubsection{Experiments in Real Datasets}
\label{sec.real_data}

We evaluate the performance of \textsc{Macomss} using EHR data collected from the Duke University Health System (DUHS), accessed through the Duke Clinical Research Datamart (CRDM) \citep{hurst2021development}. The dataset integrates EHR data from three hospitals, each treated as an isolated site: Duke Raleigh Hospital (DRAH) as site 1, Duke Regional Hospital (DRH) as site 2, and Duke University Hospital (DUH) as site 3.
Our study focuses on all emergency department (ED) visits and records clinical features of patients across three sites in 2019. To mimic the missing mechanism present in integrated analysis, we first introduce structured and sporadic missingness to the clinical data from different sources under various settings. After that, we apply imputation methods to fill in the missing values and employ the imputed data to predict whether a visit will result in inpatient admission—a classification task on the integrated EHR datasets.

For the classification, the predictors include four categories: (1) Demographics information: age and sex; (2) Vital signs: pulse (beats/min), systolic blood pressure (SBP; mm Hg), diastolic blood pressure (DBP; mm Hg), oxygen saturation (SpO$_2$; \%), temperature ($^{\circ}$F), respiration (times/min), and acuity level; (3) Comorbidities: indicators of local tumor, metastatic tumor, diabetes with complications, diabetes without complications, and renal disease, for a patient. Comorbidities are defined based on the ICD-10-CM Diagnosis Code; (4) PheCodes: in addition to known comorbidities, all other ICD-10-CM Diagnosis Codes were aggregated into PheCodes to represent more general diagnoses using the ICD-to-PheCode mapping from PheWAS catalog (\href{https://phewascatalog.org/phecodes}{https://phewascatalog.org/phecodes}). We utilized one-digit level PheCodes in the analysis. Given the low incidence rate of the response outcome, we down-sample the data by one-third for cases with outcome 0 at each site. For the final cohort, we include only samples without any initial missingness to facilitate the design of different missing data mechanisms and evaluate imputation performance accurately.

{
To implement our experiment, we collect 438 predictors from 126,579 visits, coupled with observed indicators of whether a visit will result in inpatient admission as responses.}
To generate target datasets, we first sample 400 visits and $p$ predictors from the three sources of collected data, forming a $400 \times p$ matrix, where $p$ will vary across different values.
For each visit, we always include the predictors related to demographics information, vital signs, and comorbidities, including a total of 13 features. 
The remaining $(p-13)$ features are randomly sampled from the predictors in PheCodes.
{Accordingly, we perform a log transformation and standardization on the $400 \times p$ matrix, and divide the 400 visits (including both the predictors and inpatient admission for the visit) into ten folds.}
We use nine folds as the training set and check the accuracy of classification on the remaining testing fold. 
We then output the averaged cross-validated classification accuracy for the 400 visits. 
To ensure the generalizability of our results, we repeat the above process multiple times for other randomly sampled 400 visits and $p$ predictors.

{
In the training dataset above, we introduce missing values into the \( 360 \times p \) matrix \( X \) using a sporadic missing mechanism. To be more general, we consider a non–rank-one missing matrix to explore more complicated sporadic missing mechanisms. 
Specifically, the missing probability matrix \( \Theta \) is constructed as
$\Theta = \sum_{m=1}^4 \lambda_m \alpha_m \beta_m^\top + \varepsilon_{\Theta},$
where \( \alpha_m \in \mathbb{R}^{360} \), \( \beta_m \in \mathbb{R}^p \), \( \lambda_m = 0.5 ^ {3(m-1)} \), and \( \varepsilon_{\Theta} \in \mathbb{R}^{360 \times p} \) is a matrix containing mean-zero Gaussian noise variables with standard deviation $\frac{1}{20} \left\| \sum_{m=1}^4 \lambda_m \alpha_m \beta_m^\top \right\| / \sqrt{360p}.$
Each element of \( \alpha_m \) and \( \beta_m \) is independently drawn from \( 1 - 0.2 \cdot \text{Unif}[0, 1] \).
Given the matrix \( \Theta \), we then generate the label \( M_{ij} \) by \( M_{ij} \sim \text{Bernoulli}(\Theta_{ij}) \), where \( \Theta_{ij} \) is the \( (i,j) \)-th element in \( \Theta \). If \( M_{ij} = 0 \), we set the \( (i,j) \)-th element of \( X \) as missing.
}

Additionally, we introduce structured missingness by removing observed values in $l\%$ of the rows and columns. 
The missing rows are randomly sampled from the 360 visits, while the missing columns always include important predictors related to emergency hospital admission, including demographic information, vital signs, and comorbidities \citep{wallace2014risk, lucke2018early, brink2022predicting}. The remaining missing columns are randomly sampled from the predictors of PheCodes.
We evaluate different values of $p$ and $l$ to examine cases with varying numbers of clinical features and missing proportions.

We apply \textsc{Macomss} to impute the incomplete matrix. 
To evaluate accuracies, we utilize the imputed matrices to perform logistic regressions on their responses, as described in Section~\ref{sec: DST}. 
We then output the completion errors, defined by the normalized mean squared errors (NMSEs) between the imputed values and their true observed values, with their corresponding AUC values calculated from the testing dataset, as defined in Section~\ref{sec: DST}.

\section{Results}\label{sec: resutt}

\subsection{Results for Recovery Accuracy}

\begin{figure}[ht!]
	\begin{center}
        \textbf{(A)} \\[5pt] 
		\includegraphics[height = 2.2in,width=2.7in]{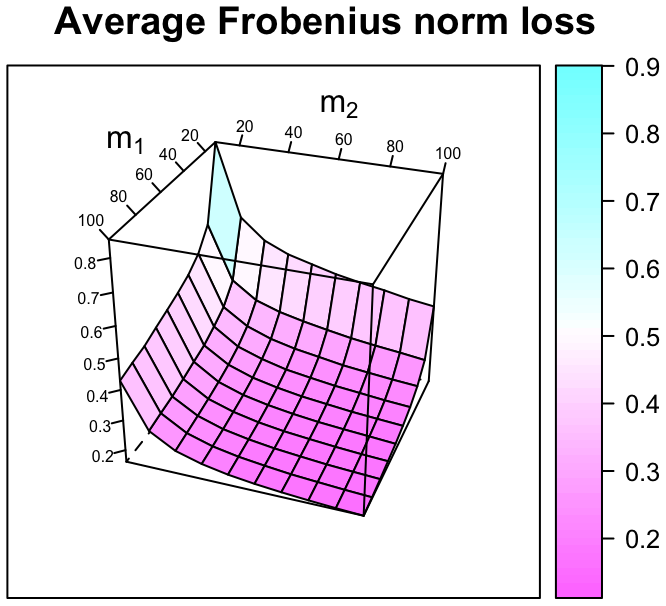}\quad  
        \includegraphics[height = 2.2in,width=2.7in]{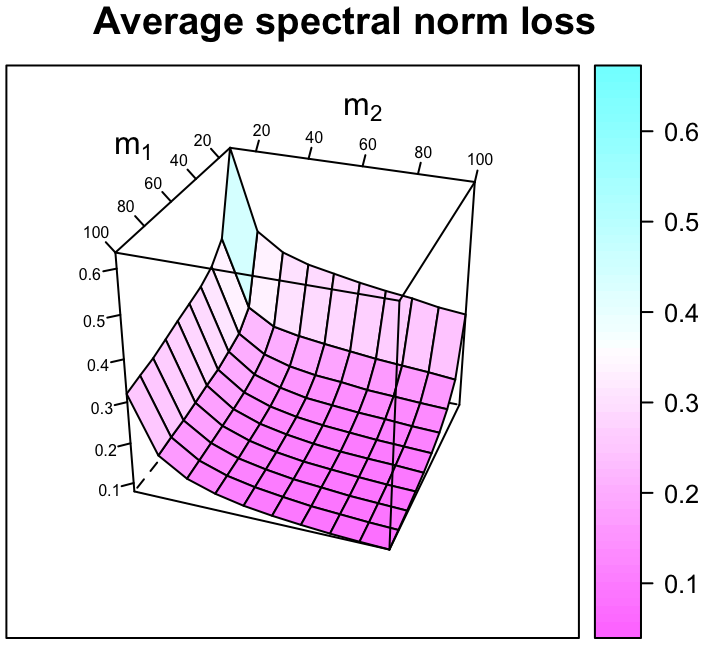}
	\end{center}
    \begin{center}
        \textbf{(B)} \\[5pt] 
		\includegraphics[height = 2.2in,width=2.7in]{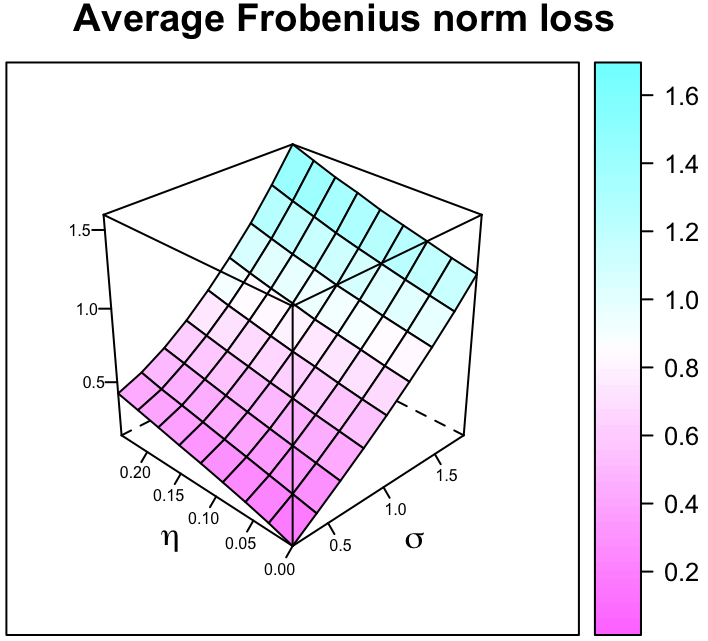}\quad  
        \includegraphics[height = 2.2in,width=2.7in]{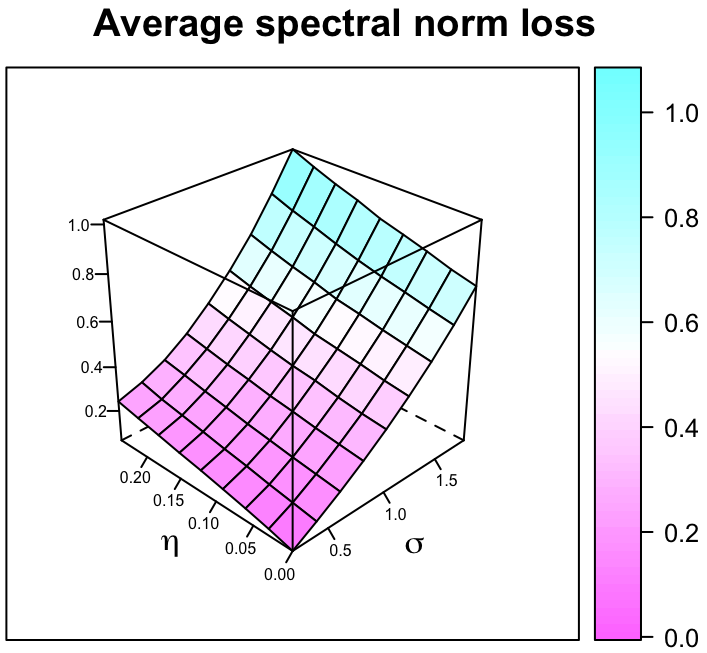}
		\caption{(A): Average Frobenius and spectral norm loss for \textsc{Macomss} with varying number of observable rows and columns: $m_1, m_2\in [10, 100]$. (B): Average Frobenius and spectral norm loss for \textsc{Macomss} for varying $\Theta$ and $\sigma$.}
		\label{fig:simu-setting2}
	\end{center}
\end{figure}

Under the simulation settings 1 - 2 in Section~\ref{sec:recorery}, we present the average Frobenius and spectral norm loss of recovery in Figure~\ref{fig:simu-setting2}, where we vary the values of $m_1$, $m_2$, $\eta$, and $\sigma$ to examine the theoretical properties and empirical performance of \textsc{Macomss}. 

By Figures~\ref{fig:simu-setting2}, it can be seen that as $m_1, m_2$ grow or $\sigma,\eta$ decrease, i.e., more rows and columns of $A$ are available or the entries of $A$ are observed with a lower noise and fewer sporadic missing values, we achieve better recovery performance using \textsc{Macomss}. These results align with the theoretical analysis in Supplementary Materials, demonstrating the stable performance of the proposed algorithm across all values of $m_1$, $m_2$, $\sigma$, and $\eta$. 

{
In Figures~\ref{fig:simu-setting3}--\ref{fig:simu-setting4} of the Supplementary Materials, we further examine the performance of \textsc{Macomss} for non–low-rank and Poisson count matrices. The results show that our method is effective for both nearly low-rank and Poisson count matrix cases, demonstrating the applicability of \textsc{Macomss}.
}

\subsection{Results for Downstream Tasks}
In this subsection, we examine the performance of \textsc{Macomss} for the downstream classification tasks under scenarios 1 - 2 in Section~\ref{sec: DST}, compared with the existing imputation methods listed in that subsection. 

\begin{figure}[h]
    \begin{center}
    \includegraphics[scale = 0.7]{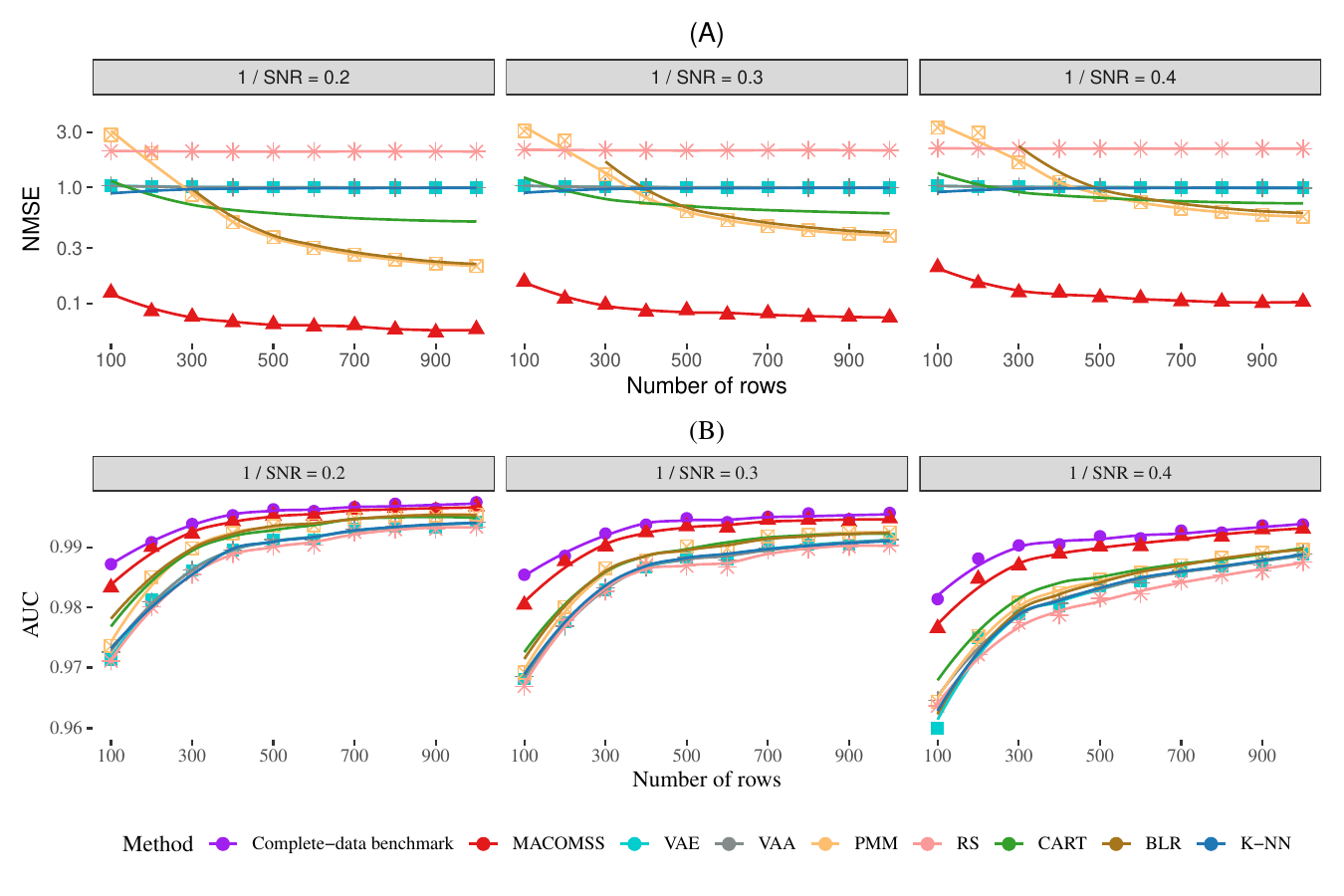}
    \caption{The $\text{NMSE}$ (A) and the {$\text{AUC}$ (for simulated binary outcomes) (B)} for scenario 1 for different numbers of rows from different methods.}
    \label{fig:performance-compare-1}
    \end{center}
\end{figure}

We first consider scenario 1, with the number of rows, \( n \), ranging from 100 to 1000, and the number of
columns, $p$, fixed at 70. The corresponding NMSEs and AUCs are shown in Figure \ref{fig:performance-compare-1}. In Figure~\ref{fig:performance-compare-1}(A), we observe that \textsc{Macomss} consistently outperforms other methods, achieving smaller NMSEs across all values of $n$ and SNR. 
Moreover, the performance gap between \textsc{Macomss} and the other competing methods widens as SNR decreases, since the latter do not denoise the observed entries.
As a result, our method yields more high AUC values, indicating more accurate estimates of regression coefficients and better classification accuracy for the downstream classification task. These advantages become more pronounced as the noise level increases of the data (see Panel (B) of Figure \ref{fig:performance-compare-1}). 

\begin{figure}[h]
    \begin{center}
        \includegraphics[scale = 0.5]{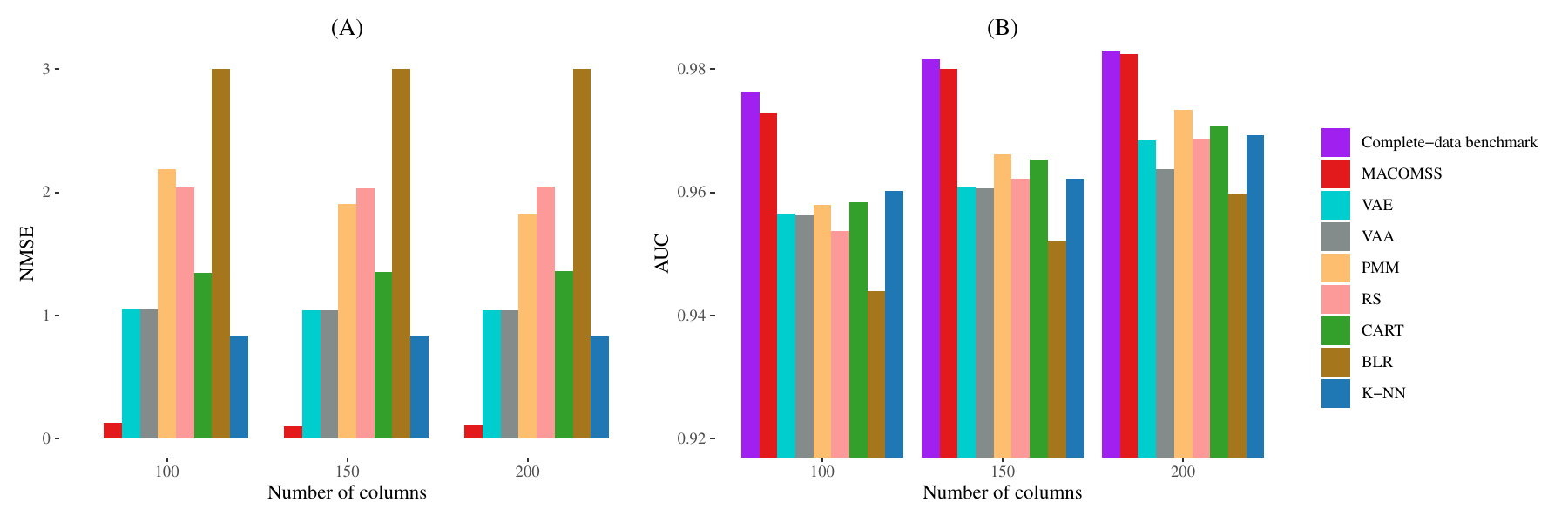}
    \caption{The $\text{NMSE}$ (A) and {$\text{AUC}$ (for simulated binary outcomes) (B)} for scenario 2 for different numbers of columns from different methods. We set the SNR to 5 for data generation.
    }
    \label{fig:performance-compare-2}
    \end{center}
\end{figure}

Moreover, we consider the high-dimensional setting under scenario 2, where the number of rows, \( n \), is fixed at 70, and the number of columns, \( p \), varies from 100 to 200. The corresponding NMSEs and AUCs are presented in Figure \ref{fig:performance-compare-2}. In these cases, we similarly observe that \textsc{Macomss} outperforms other methods in terms of NMSE and AUC.

In both scenarios 1 and 2, we find that the AUCs of \textsc{Macomss} are close to that of the complete-data case (Panel (B) in Figures \ref{fig:performance-compare-1} - \ref{fig:performance-compare-2}), whether we have a large missing block with the increased number of rows (patients) or columns (clinical features). 
These results indicate that \textsc{Macomss} achieves nearly optimal performance in the predictive task using the imputed matrix, owing to its consideration of flexible sporadic and structural missingness mechanisms for data imputation.

\subsection{Results for Real EHR Datasets}

\begin{figure}[h]
    \begin{center}
    \includegraphics[scale = 0.8]{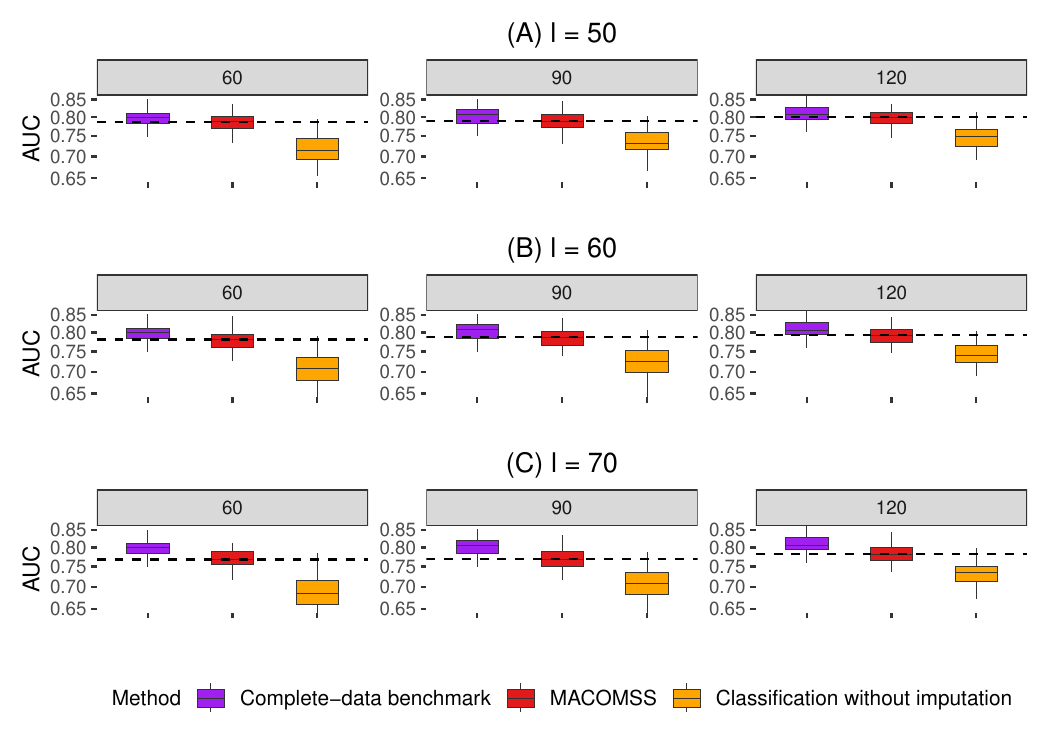}
    \caption{The {$\text{AUC}$ (for admission prediction)} from 50-times experiments with the missing proportions, $l\%$ (main title), ranging from 50 to 70, and the number of features (sub-title), $p$, ranging from 60 to 120. The dashed horizontal lines indicate the median values of AUC from \textsc{Macomss}.}
    \label{fig:performance-compare-EHR-extra}
    \end{center}
\end{figure}

{
Before performing analysis, we examine the validity of the low-rank assumption for the EHR dataset in Part~\ref{sec: low-rank} of the Supplementary Materials. Our results show that the data matrix are nearly low-rank, which fulfills the requirement of \textsc{Macomss}.}
Using these EHR datasets, we conduct experiments with different values of \( p \) and \( l \) under the setting described in Section~\ref{sec.real_data}. In these experiments, we randomly sample 400 patients and \( p \) clinical features with a missing block controlled by \( l \), repeated 50 times. We apply \textsc{Macomss} to impute these matrices for downstream classification tasks, where the classification accuracy is measured by AUC in testing sets.
For comparison, we compute the AUC for logistic regressions using the complete predictor matrix, serving as the complete-data method for classification performance. To examine the impact of block missingness, we also report the AUC obtained by removing clinical features with structural missingness -- a conventional approach that disregards structural patterns in downstream tasks. These AUC values are shown in Figure~\ref{fig:performance-compare-EHR-extra}.

We observe that \textsc{Macomss} significantly outperforms classification without imputation, as evidenced by the AUC values in Figure~\ref{fig:performance-compare-EHR-extra}. Notably, this recovery achieves a prediction accuracy close to the complete-data method, demonstrating the generalizability of \textsc{Macomss} for clinical prediction tasks.

\begin{figure}[ht!]
    \centering
    \includegraphics[width=\linewidth]{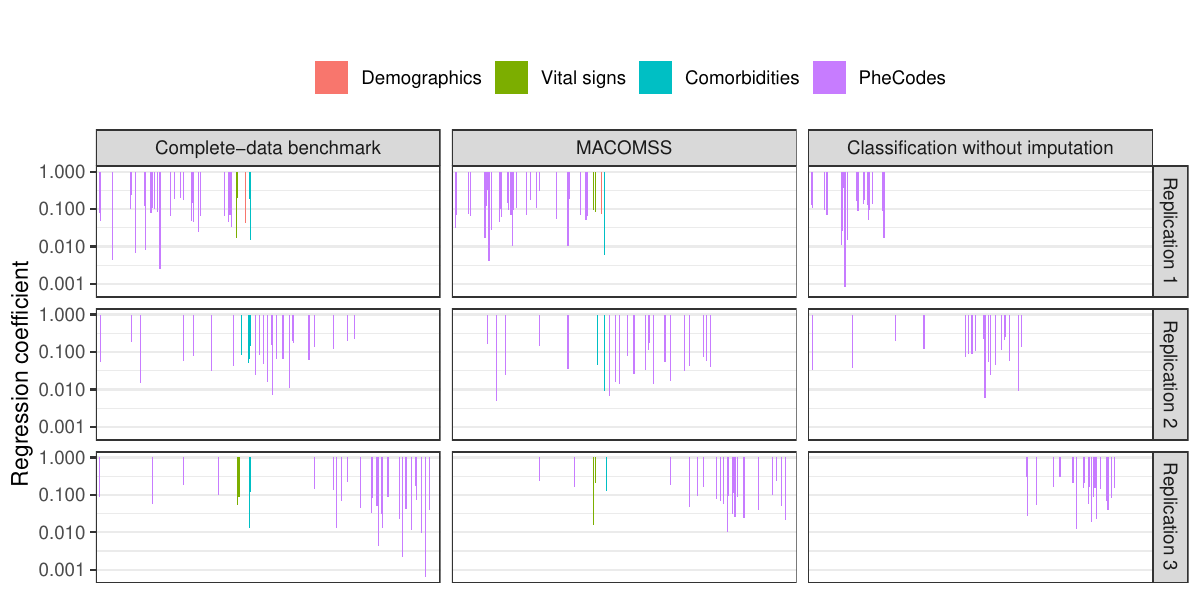}
    \caption{Estimated logistic regression coefficients from three randomly selected replication experiments under the complete-data benchmark, \textsc{Macomss}, and classification without imputation.}
    \label{fig:RC}
\end{figure}

{
In Figure~\ref{fig:RC}, we illustrate the coefficients from logistic regression models using the above methods, estimated from three randomly selected replication experiments in Figure~\ref{fig:performance-compare-EHR-extra} with $p = 120$ and $l = 50$. We find that \textsc{Macomss} identifies clinical features in demographic information, vital signs, and comorbidities—similar to the complete-data benchmark. These features are usually important predictors for emergency hospital admission \citep{wallace2014risk, lucke2018early, brink2022predicting}, but they are ignored when no imputation is performed. These results highlight the necessity of imputing block missingness in the predictor matrix prior to classification, as the imputation may recover important information crucial for predicting patient admission.
}

\begin{figure}[h]
    \begin{center}
    \includegraphics[width=\textwidth, height=8cm]{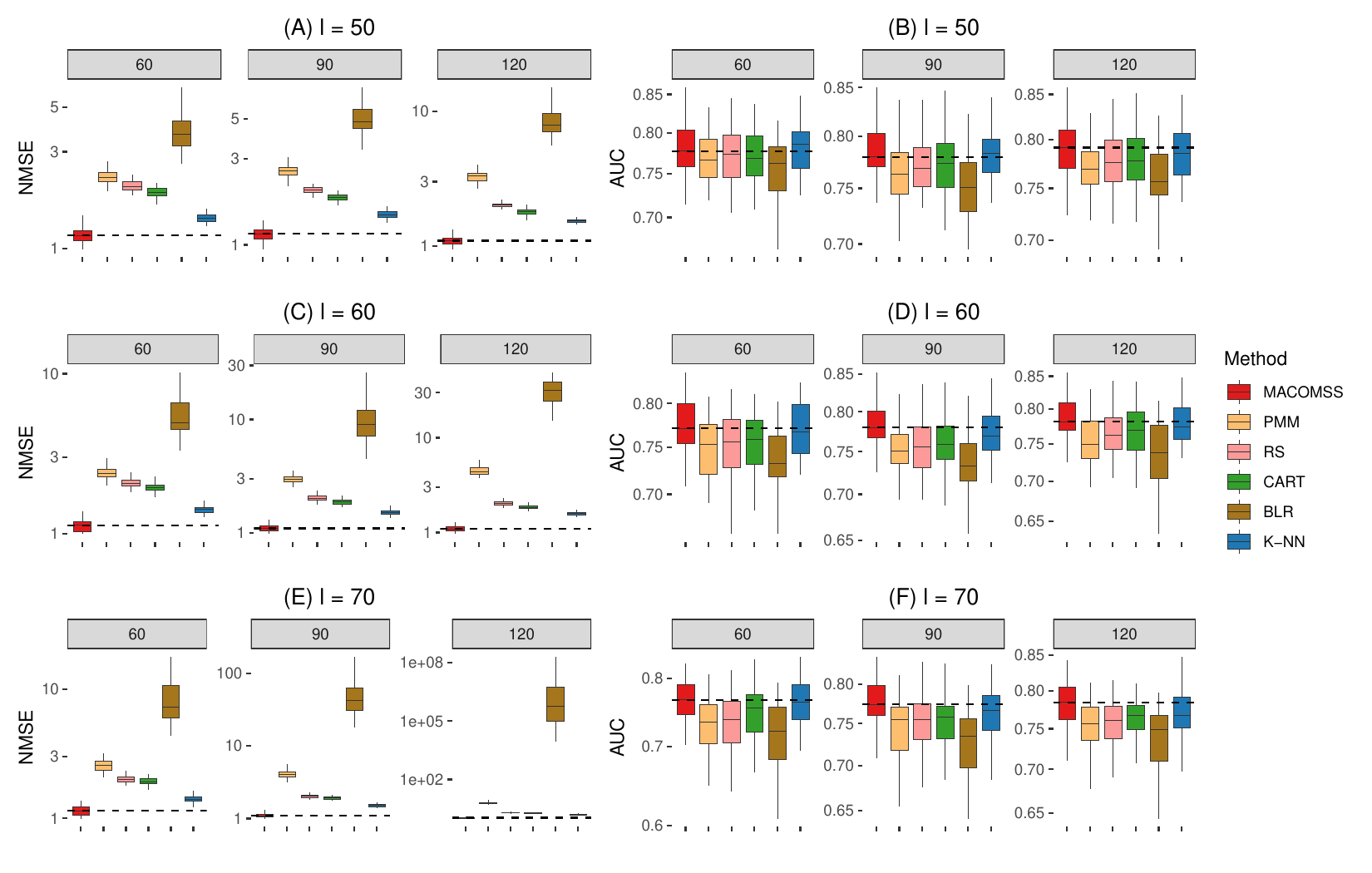}
    \caption{The NMSE ((A), (C), and (E)) and AUC {for admission prediction} ((B), (D), and (F)) with the missing proportions, $l\%$ (main title), ranging from 50 to 70, and the number of features (sub-title), $p$, ranging from 60 to 120. The dashed horizontal lines indicate the median values of NMSE or AUC from \textsc{Macomss}.}
    \label{fig:performance-compare-EHR-2}
    \end{center}
\end{figure}

{
We further compare \textsc{Macomss} with PMM, RS, CART, BLR, and K-NN listed in Section~\ref{sec: DST}. We evaluate the imputation accuracy using NMSE, along with the corresponding AUC to assess the performance of downstream predictive tasks. The NMSEs and AUCs of different methods across 50 repeated experiments are presented in Figure~\ref{fig:performance-compare-EHR-2}.
We observe that \textsc{Macomss} mostly outperforms other methods in terms of NMSE (Panel~(A), (C), and (E) in Figure~\ref{fig:performance-compare-EHR-2}). This satisfactory imputation performance of \textsc{Macomss} leads to superior prediction accuracy for downstream classification tasks, as evidenced by the AUC results in Panels~(B), (D), and (F) of Figure~\ref{fig:performance-compare-EHR-2}. 
}

\begin{figure}[h]
    \begin{center}
    \includegraphics[scale = 0.6]{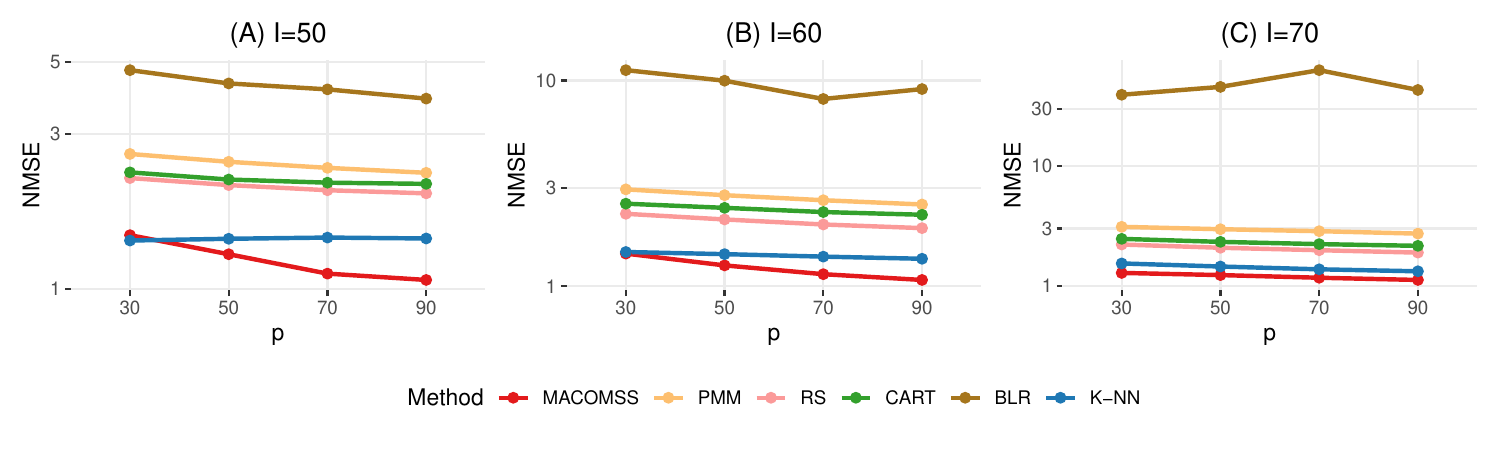}
    \caption{The averaged NMSE is computed over 50 repeated experiments with varying missing proportions $l\%$ and $p=120$. For each experiment, we randomly partition the patients into three groups and generate a new dataset by adding Gaussian noise to the original dataset according to these groups. For each clinical feature in the $k$-th group ($k=1,2,3$), the mean and standard deviation of the added noise are set to $(100 - (k-2)h)\%$ of the feature’s mean and $10\%$ of its standard deviation, respectively. A larger $h\%$ corresponds to a higher level of population bias and heterogeneity.}
    \label{fig:performance-compare-EHR-3}
    \end{center}
\end{figure}

{
We also compute the runtime and memory usage benchmarks for the above methods using the DUHS dataset. For dataset of dimension $360 \times p$, with $p \in \{60, 90, 120\}$, we report the averaged runtime and memory consumption for each method in Table~\ref{tab:runtime_memory}, evaluated on a server with 2.10~GHz CPU cores and 208~GB of RAM. These results show that \textsc{Macomss} achieves favorable computational efficiency compared to other methods.
}

\begin{table}[h]
\centering
\caption{Average runtime (in seconds) and memory usage (in MB) over 50 repeated experiments with $p \in \{60, 90, 120\}$ for different methods on the DUHS dataset.}
\label{tab:runtime_memory}
\begin{tabular}{ccccccc}
\hline
 & \textsc{Macomss} & PMM & RS & CART & BLR & K-NN \\ 
\hline
Time & 0.072 & 4.166 & 4.173 & 3.484 & 7.771 & 96.707 \\ 
  Memory & 17.631 & 635.871 & 623.902 & 464.169 & 812.366 & 22859.319 \\
\hline
\end{tabular}
\end{table}

We perform additional experiments to compare the performances between \textsc{Macomss} and previous methods by increasing the heterogeneity level in the data. {To this end, we randomly divide the patients in the clinical dataset into three groups and generate a new dataset by adding group-specific noise, increasing the population bias and heterogeneity level in the data.}
Subsequently, we apply \textsc{Macomss} and other imputation methods to the new dataset and replicate the previous experiments 50 times. 
The averaged NMSE under different heterogeneity levels are presented in Figure~\ref{fig:performance-compare-EHR-3}. We observe that the NMSE of \textsc{Macomss} becomes smaller compared to that of the other as the heterogeneity level increases.  {These results suggest that when significant bias and heterogeneity is present in different data sources, existing methods may no longer provide satisfactory imputation for structured and sporadic missingness. Whereas \textsc{Macomss}, in general, can achieve superior performance for this setting, which highlights its applicability for integrative analysis of EHR datasets with potential heterogeneity.}

\section{Discussion}
\label{sec.discussion}

This article proposes a novel matrix completion method named \textsc{Macomss} for imputing datasets with structured and sporadic missingness, effectively addressing key missingness occurring in the integrated analysis of EHR data. Our approach preserves critical information that may be lost due to the intricate healthcare delivery and data collection systems and ensures that downstream clinical and public health analyses remain robust and more generalizable to a larger population.
With the theoretical guarantees, simulation validation, and real data analysis, our method demonstrates superiority in data imputation during integrated analyses, particularly in multi-source EHR studies where traditional imputation methods struggle. Its resilience to data heterogeneity further ensures that population-level health patterns can be studied with greater precision, facilitating better clinical decision-making and more effective public health interventions.

{
Our work has several limitations that point to directions for future research. 
First, \textsc{Macomss} is designed to treat different data types in a uniform way and does not explicitly accommodate binary or mixed-type variables; in such cases,  techniques such as logistic PCA~\citep{lee2010sparse} or mixed-type factor models \citep{liu2023generalized} can be considered either as preprocessing steps or in conjunction with our approach. 
Second, the current formulation does not exploit temporal smoothness or time-indexed missing patterns that may be present in EHR datasets, which could lead to a loss of estimation efficiency in longitudinal or time series settings. 
To address this, one could extend \textsc{Macomss} with time-based regularization or penalty terms on the singular value decomposition, similar to recent works~\citep{han2024guaranteed,tan2024functional}. 
Third, while we address missing-at-random (MAR)-like missingness, missing-not-at-random (MNAR) remains a challenging scenario for EHR data imputation. 
In Part~\ref{asup_eval} of the Supplementary Materials, we provide a sensitivity analysis to explore robustness under the MNAR mechanism, but further methodological development is needed for principled handling of such patterns.
}

{
Overall, \textsc{Macomss} inherits challenges common to large-scale, heterogeneous EHR datasets, where diverse data sources, variable quality, and complex missingness patterns can complicate integrated analysis. 
To apply \textsc{Macomss} or other imputation approaches effectively, we recommend a thorough examination of missingness patterns and an assessment of potential population biases prior to EHR analysis, along with applying appropriate transformations for variable types when needed. 
In parallel, it is essential to validate imputations through downstream predictive performance and evaluations of clinical plausibility, ideally with guidance from domain experts to ensure both statistical rigor and real-world applicability.
}

\section{Conclusion}
\textsc{Macomss} provides an easy-to-implement and theoretically guaranteed approach for imputing and denoising a matrix integrated from multiple EHR data sources. 
Our approach effectively leverages the complicated missing mechanisms during imputation, addressing key challenges where missingness may occur randomly, structurally, and heterogeneously in the data integration processes.
These advantages highlight the potential of our method for accurate downstream clinical prediction and precise clinical insights in population health.

\section*{Data and Code Availability}
The codes to implement \textsc{Macomss} are publicly available at \href{https://github.com/Tan-jianbin/Macomss}{https://github.com/Tan-jianbin/Macomss}.

\section*{CRediT Authorship Contribution Statement}

{\bf Jianbin Tan}: Writing – review \& editing, Writing – original draft, Software, Investigation, Formal analysis. Yan Zhang:  Writing – review \& editing, Investigation, Formal analysis, Software, Data curation; {\bf Chuan Hong}: Data curation, Supervision, Writing – review \& editing. {\bf T. Tony Cai}: Writing – review \& editing, Conceptualization, Supervision. {\bf Tianxi Cai}: Writing – review \& editing, Conceptualization, Supervision. {\bf Anru R. Zhang}: Methodology, Writing – review \& editing, Writing – original draft, Supervision, Software, Conceptualization.

\section*{Declaration of generative AI and AI-assisted technologies in the writing process}
During the preparation of this work the authors used ChatGPT in order to improve the readability and language of the manuscript. After using this tool/service, the authors reviewed and edited the content as needed and took full responsibility for the content of the published article.

\section*{Funding Information}

This work was supported in part by the National Institutes of Health Grant R01HL169347.

\bibliographystyle{apalike}
\bibliography{reference.bib}


\clearpage
\newpage
\appendix

\section{Theoretical Framework of \textsc{Macomss}}\label{sec.theory}

Before getting into details, we define addition notations relating to theoretical analysis. Let $\mathbb{I}(\cdot)$ denote the indicator function, and $I_p$ be the identity matrix in $\mathbb{R}^{p \times p}$.
We also denote $e_i^{(p)}$, the $i$-th canonical basis in $p$-dimensional space, as the $p$-dimensional vector with $i$-th entry as 1 and others as zero. For \( A \in \mathbb{R}^{p_1 \times p_2} \), we use \( A_{[\Omega_1, \Omega_2]} \) to represent the sub-matrix of \( A \) with row indices \( \Omega_1 \) and column indices \( \Omega_2 \). For convenience, we denote \( a:b = \{a, \ldots, b\} \) and ``:" as the entire index set. Therefore, \( A_{[:, 1:r]} \) stands for the first \( r \) columns of \( A \), while \( A_{[1:m_1, 1:m_2]} \) represents the first \( m_1 \) rows and first \( m_2 \) columns of \( A \). 
Define the sub-Gaussian norm for any random variable $Y$ as $\|Y\|_{\phi_2} := \sup_{q\geq 1}\left(\E|Y|^q\right)^{1/q}/\sqrt{q}$, see, e.g., \cite{vershynin2010introduction}). 

When \( A \in \mathbb{R}^{p_1 \times p_2} \), we denote \( \sigma_1(A) \geq \sigma_2(A) \geq \cdots \geq \sigma_{p_1 \wedge p_2}(A) \) as the set of all singular values of \( A \) in descending order. In particular, let \( \sigma_{\min}(A) = \sigma_{p_1 \wedge p_2}(A) \) and \( \sigma_{\max}(A) = \sigma_1(A) \) represent the minimum and maximum singular values of \( A \), respectively. Note that \( \sigma_1(A) = \|A\| \), the spectral norm of \( A \), and \( \sqrt{\sum_{r=1}^{p_1 \wedge p_2}\sigma_r^2(A)} = \|A\|_F \), the Frobenius norm of \( A \). In the analysis, orthogonal matrices will play a crucial role, we thus denote \( \mathbb{O}_{p, r} = \{U \in \mathbb{R}^{p \times r}: U^\top U = I_r\} \), the set of all \( p \)-by-\( r \) orthogonal matrices. 
Finally, let \( C, c, C_1, C_2, \ldots \) be generic symbols for constants, whose actual values may vary from line to line.

\begin{algorithm}
	\caption{Matrix Completion with Noise and Structured and Sporadic Missingness}
\footnotesize
	\begin{algorithmic}[1]
		\State Input: $Y\in \mathbb{R}^{p_1\times p_2}$ with missing indicator $M\in\{0, 1\}^{p_1\times p_2}$.
		\State Estimate $\hat{\Theta} = (\Theta_{ij})_{1\leq i \leq p_1, 1\leq j\leq p_2}$ as
		\begin{equation*}
		\begin{split}
		& \hat{\Theta}^{\rm (col)} = (\hat{\Theta}_{ij}^{\rm (col)})\in \mathbb{R}^{p_1\times m_2}, \quad \hat{\Theta}^{\rm (col)}_{ij} = \left(\sum_{j'=1}^{m_2} M_{ij'} \right) \left(\sum_{i'=1}^{p_1} M_{i'j}\right) / \left(\sum_{i'=1}^{p_1}\sum_{j'=1}^{m_2} M_{i'j'}\right),\\ 
		& \hat{\Theta}^{\rm (row)} = (\hat{\Theta}_{ij}^{\rm (row)})\in \mathbb{R}^{m_1\times p_2}, \quad \hat{\Theta}^{\rm (row)}_{ij} = \left(\sum_{j'=1}^{p_2} M_{ij'} \right) \left(\sum_{i'=1}^{m_1} M_{i'j}\right) / \left(\sum_{i'=1}^{m_1}\sum_{j'=1}^{p_2} M_{i'j'}\right).
		\end{split}
		\end{equation*}
	\begin{equation*}
\begin{split}
& \hat{\Theta}_{ij} = \left\{\begin{array}{ll}
\left(\frac{1}{2}(\hat{\Theta}^{\rm (col)}_{ij})^{-1} + \frac{1}{2}(\hat{\Theta}^{\rm (row)}_{ij})^{-1}\right)^{-1}, & i \in[1:m_1], j\in [1:m_2]\\
\hat{\Theta}_{ij}^{\rm (row)}, & i\in [1:m_1], j\in [(m_2+1):p_2]\\
\hat{\Theta}_{ij}^{\rm (col)}, & i\in [(m_1+1):p_1], j\in [1:m_2];\\
0 & \text{otherwise}.
\end{array}\right.
\end{split}
\end{equation*}
$$\hat{\Theta}_{(11)} = (\hat{\Theta}_{ij})_{\substack{1\leq i \leq m_1\\1\leq j \leq m_2}}, \quad\hat{\Theta}_{(12)} = (\hat{\Theta}_{ij})_{\substack{1\leq i \leq m_1\\m_2+1\leq j \leq p_2}},\quad \hat{\Theta}_{(21)} = (\hat{\Theta}_{ij})_{\substack{m_1+1 \leq i \leq p_1\\1\leq j \leq m_2}}.$$
		\State Set $Y_{ij} = 0$, whenever $M_{ij}= 0$. Calculate
		$$\tilde{Y} \in \mathbb{R}^{p_1\times p_2}, \quad \tilde{Y}_{ij} = Y_{ij}/\hat{\Theta}_{ij}, \quad 1\leq i \leq p_1, 1\leq j \leq p_2.$$
		\State Calculate the SVDs for 
		$$Y^{\rm (col)} = \begin{bmatrix}
\frac{(p_1-2m_1)\wedge 0}{p_1}\cdot \tilde{Y}_{(11)} \\
\tilde{Y}_{(21)}
\end{bmatrix}, \quad \text{and}\quad Y^{\rm (row)} = \left[\frac{(p_2-2m_2)\wedge 0}{p_2}\cdot \tilde{Y}_{(11)}, ~ \tilde{Y}_{(12)}\right],$$ 
		Let $V^{\rm (col)}$ and $U^{\rm (row)}$ be the right singular vectors of $\tilde{Y}^{\rm (col)}$ and the left singular vectors of $\tilde{Y}^{\rm (row)}$, respectively. Then calculate
		\begin{equation*}
		B_{(11)} = \{U^{\rm (row)}\}^\top\tilde{Y}_{(11)} V^{\rm (col)},\quad B_{(12)} = \{U^{\rm (row)}\}^\top\tilde{Y}_{(12)}, \quad B_{(21)} = \tilde{Y}_{(21)} V^{\rm (col)}.
		\end{equation*}
		\For {$s = r_0: -1: 1$}
		\State {\bf if} $B_{(11), [1:s, 1:s]}$ is not singular and 
		$$\left\|B_{21, [:, 1:\hat{r}]}B_{11, [1:\hat{r}, 1:\hat{r}]}^{-1}\right\| \leq 2\sqrt{p_1/m_1}\quad \text{or}\quad \left\|B_{11, [1:\hat{r}, 1:\hat{r}]}^{-1} B_{12, [1:\hat{r}]}\right\| \leq 2\sqrt{p_2/m_2}$$ \quad~ {\bf then}
		\State \quad\quad $\hat r = s$; {\bf break} from the loop;
		\State {\bf end if}
		\EndFor		
		\State{\bf If} $\hat{r}$ is still unassigned {\bf then} $\hat{r}_t = 0$.
		\State We calculate the final estimator $\hat{A}$, where
		\begin{equation*}
		\begin{split}
		& \hat{A}_{(11)} = U^{\rm (row)}_{[:, 1:\hat{r}]}B_{(11), [1:\hat{r}, 1:\hat{r}]}V^{\rm (col)\top}_{[:, 1:\hat{r}]}, \quad \hat{A}_{(12)} = U^{\rm (row)}_{[:, 1:\hat{r}]}B_{(12), [1:\hat{r}, :]},\\
		& \hat{A}_{(21)} = B_{(21), [:, 1:\hat{r}]}V^{\rm (col)\top}_{[:, 1:\hat{r}]}, \quad\quad\qquad\quad  \hat{A}_{(22)} = B_{(21), [:, 1:\hat{r}]}B_{(11), [1:\hat{r}, 1:\hat{r}]}^{-1}B_{(12), [1:\hat{r}, :]}.
		\end{split}
		\end{equation*}	\end{algorithmic}\label{al:procedure}
\end{algorithm}

We propose the procedure of \textsc{Macomss} in Algorithm~\ref{al:procedure} and analyze the theoretical performance of the proposed procedure in what follows. To begin, we state the assumptions that will be used in the theoretical analysis.

\begin{Condition}[Sub-Gaussian Observations] \label{con:sub-gaussian}
	All entries $Y_{ij}$ are independent and $\|Y_{ij}\|_{\psi_2} \leq \tau$ for some constant $\tau>0$. Here the $\|\cdot\|_{\phi_2}$ is the sub-Gaussian norm. In addition, $\sum_{i=1}^{m_1}\sum_{j=1}^{m_2} \E(Y_{ij} - A_{ij})^2 \geq \frac{1}{(m_1m_2)^c}$. 
    \end{Condition}
Condition \ref{con:sub-gaussian} characterizes that all observations are independently sub-Gaussian distributed. The commonly used identical distribution \citep{keshavan2010matrix} or bounded assumption \citep{chatterjee2015matrix} in matrix completion literature are not necessary in our analysis. In addition, we impose a mild condition on the lower bound of noise, i.e. $\sum_{i=1}^{m_1}\sum_{j=1}^{m_2}\E(Y_{ij} - A_{ij})^2 \geq \frac{1}{(mn)^c}$, which will be used only for technical purposes. 

\begin{Condition}[Low-rank and Incoherence Assumption]\label{con:low-rank-incoherence} 
	Suppose $\rank(A) = r$ and $A = U\Sigma V^\top$ is the SVD, where  $U\in \mathbb{O}_{p_1, r}$ and $V\in \mathbb{O}_{p_2, r}$ satisfy the incoherence condition, i.e. for constant $\rho>0$ the following inequality holds,
	\begin{equation*}
	\frac{p_1}{r}\max_{i}\left\|\mathbb{P}_{U}e_i^{(p_1)}\right\|_2^2 \leq \rho, \quad \frac{p_2}{r}\max_{j}\left\|\mathbb{P}_{V}e_j^{(p_2)}\right\|_2^2 \leq \rho. 
	\end{equation*}
\end{Condition}
We also assume Condition \ref{con:low-rank-incoherence}, i.e. the low-rank and incoherence assumption, holds for the matrix $A$. Especially, the incoherence condition is widely used in matrix completion literature \citep{candes2009exact,recht2011simpler,chen2015incoherence}, which guarantees that every entry of $A$ contains similar amount of information. As we will show later, the observable entries in $Y_{(11)}, Y_{(12)}$, and $Y_{(21)}$ will contain enough knowledge to infer the whole missing block of $A_{(22)}$ by assuming Condition \ref{con:low-rank-incoherence}.

\begin{Condition}[Missingness]\label{con:missingness}
The missing value indicators $M\in \mathbb{R}^{p_1\times p_2}$ are independently Bernoulli distributed: $M_{ij} \sim {\rm Bern}(\Theta_{ij})$, whenever $1\leq i \leq m_1$ or $1\leq j \leq m_2$. Here, 
\[
\theta_0 := \min \left\{\Theta_{ij}: 1\leq i\leq m_1 \text{ or } 1\leq j \leq m_2\right\} \geq Cr(m_1\wedge m_2)^{-1}\log (p_1\vee p_2)
\] 
and $\rank(\Theta) =1$.
\end{Condition}

	When there are possibly sporadic missing entries in $Y_{(12)}, Y_{(21)}$, and $Y_{(11)}$, we further assume Condition \ref{con:missingness} holds. The rank-1 assumption is widely seen in matrix completion literature (see, e.g. \cite{foygel2011learning,loh2012high}). The condition that $\theta_0 \geq Cr(m_1 \wedge m_2)^{-1}\log(p_1\vee p_2)$ is also widely seen in matrix completion literature (see, e.g. \cite{candes2010power,recht2011simpler}).

Based on these conditions, we have the following theoretical guarantees for the proposed procedure for the noisy matrix completion with structured and sporadic missingness. We particularly consider the following class of low-rank matrices,
\begin{equation}\label{eq:class-A}
\mathcal{A}_{p_1, p_2}(r,\lambda) = \left\{A \in \mathbb{R}^{p_1\times p_2}: \rank(A) = r, \sigma_r(A) \geq \lambda\right\}.
\end{equation} 

\begin{Theorem}[Upper bound]\label{th:upper_bound}
Assume $ m_1 \wedge m_2 \geq Cr\log(er), p_1 \wedge p_2 \geq m_1 \vee m_2$.  Suppose we uniformly randomly observe $m_1$ rows and $m_2$ columns from $A$ with possible noise, and there are possible missing values in $Y_{(11)}, Y_{(12)}, Y_{(21)}$. Assume Conditions \ref{con:sub-gaussian}, \ref{con:low-rank-incoherence} and \ref{con:missingness} are met, $r_0$ is selected as \eqref{eq:selection of r_0} \eqref{eq:theta_0_tau} in Supplementary Materials, then there exists a constant $C_{gap}, \eta > 0$ such that the proposed estimator in Algorithm \ref{al:procedure} yields the following rate of convergence for $\lambda = C \left(\frac{p_1p_2\log(p_1\vee p_2)}{\theta_0(m_1\wedge m_2)}\right)^{1/2}$ with $C> C_{gap}$,
\begin{equation}\label{ineq:upper_bound}
\sup_{A\in \mathcal{A}_{p_1, p_2}\left(r, \lambda\right)}\E \|\hat A - A\|^2 \leq C \frac{p_1p_2\log (p_1\vee p_2)}{\theta_{0}(m_1\wedge m_2)},
\end{equation}
\begin{equation}
\sup_{A\in \mathcal{A}_{p_1, p_2}\left(r, \lambda\right)}\E \|\hat A - A\|_F^2 \leq C \frac{p_1p_2r\log (p_1\vee p_2)}{\theta_{0}(m_1\wedge m_2)}.
\end{equation}
Moreover, when there is no sporadic missingness, i.e. $\theta_0=1$, we have the following sharper upper bound,
\begin{equation}\label{ineq:upper_bound_no_sporadic}
\sup_{A\in \mathcal{A}_{p_1, p_2}\left(r, \lambda\right)}\E \|\hat A - A\|^2 \leq C \frac{p_1p_2}{m_1\wedge m_2},
\end{equation}
\begin{equation}\label{ineq:upper_bound_no_sporadic_F}
\sup_{A\in \mathcal{A}_{p_1, p_2}\left(r, \lambda\right)}\E \|\hat A - A\|_F^2 \leq C \frac{p_1p_2r}{m_1\wedge m_2}.
\end{equation}
\end{Theorem}

Additionally, the result in Theorem \ref{th:upper_bound} is nearly rate-optimal as we have the following lower bound result.
\begin{Theorem}[Lower bound]\label{th:lower_bound} Suppose $m_1\wedge m_2 \geq Cr\log(er); p_1 \wedge p_2 \geq 2(m_1 \vee m_2)$, $\Omega_1\subseteq \{1,\ldots, p_1\}$, $\Omega_2\subseteq\{1,\ldots, p_2\}$, such that $|\Omega_1| = m_1, |\Omega_2| = m_2$. $Y_{(11)}$, $Y_{(12)}$, and $Y_{(21)}$ are defined as \eqref{eq:Y_block}. When $\lambda = c_{gap}\left(\frac{p_1p_2}{\theta_0(m_1\wedge m_2)}\right)^{1/2}$, we have the lower bound result in the class of low-rank matrices $\tilde{\mathcal{A}}_{p_1,p_2}(r,\lambda)$:
	\begin{equation}\label{ineq:lower_bound}
	\inf_{\hat{A}} \sup_{\substack{A\in \tilde{\mathcal{A}}_{p_1, p_2}(r,\lambda)}} \E \|\hat{A} - A\|^2 \geq c\frac{p_1p_2}{\theta_{0}(m_1\wedge m_2)},\quad 
	\inf_{\hat{A}} \sup_{\substack{A\in \tilde{\mathcal{A}}_{p_1, p_2}(r,\lambda)}} \E \|\hat{A} - A\|_F^2 \geq c\frac{p_1p_2r}{\theta_{0}(m_1\wedge m_2)}.
	\end{equation}	
where $\tilde{\mathcal{A}}_{p_1,p_2}(r,\lambda)$ is the class of matrices $\mathcal{A}_{p_1,p_2}(r,\lambda)$ satisfying Conditions \ref{con:sub-gaussian}, \ref{con:low-rank-incoherence}, and \ref{con:missingness}	
	
	Moreover, there exists uniform small constant $c_{gap}, c>0$, such that the following lower bound holds for the class of low-rank matrices,
	\begin{equation}\label{ineq:lower_bound_2}
	\inf_{\hat{A}} \sup_{A\in \tilde{\mathcal{A}}_{p_1, p_2}(r,\lambda)} \E \frac{\|\hat{A} - A\|^2}{\|A\|} \geq c,\quad 
	\inf_{\hat{A}} \sup_{ A\in \tilde{\mathcal{A}}_{p_1, p_2}(r,\lambda)} \E \frac{\|\hat{A} - A\|_F^2}{\|A\|_F} \geq c.
	\end{equation}

\end{Theorem}

\begin{remark}
Theorem \ref{th:lower_bound} provides critical lower bound support for Theorem \ref{th:upper_bound} in two aspects: 
first, inequality \eqref{ineq:lower_bound} demonstrates that the proposed procedure in Algorithm \ref{al:procedure} achieves rate-optimality within the given class of low-rank matrices. This confirms the statistical optimality of the method in recovering structured and sporadically missing data. Second, inequality \eqref{ineq:lower_bound_2} underscores the importance of the singular value gap condition on $\sigma_r(A)$ in the theoretical analysis for the upper bound. Failure to satisfy this condition may result in inconsistent estimates, highlighting its necessity for reliable matrix completion.
\end{remark}

\begin{remark}
It is worth noting that the proposed procedure becomes a more general version of Singular Value Thresholding (SVT) for matrix completion \citep{chatterjee2015matrix} when there is no structural missingness (i.e., $m_1 = p_1$ and $m_2 = p_2$). Besides, our method reduces to structured matrix completion \citep{cai2016structured} when there is no sporadic missingness (i.e., $\Theta_{ij} = 1$), and to hard singular value thresholding (HSVT) for matrix denoising \citep{candes2013unbiased, donoho2014minimax, gavish2014optimal} when there are no missing values at all (i.e., $m_1 = p_1$, $m_2 = p_2$, and $\Theta_{ij} = 1$).
We briefly compare them in theoretical performance here.
	\begin{itemize}
		\item When there are no sporadic missingness, i.e. $\theta_0 = 1$, the proposed method and theoretical analysis yields the following upper bound
	\[
	\E \|\hat{A} - A\|^2 \leq \frac{Cp_1p_2}{m_1\wedge m_2}.
	\]
		This result outperforms the one in Corollary 1 in \cite{cai2016structured}, which only yields $\E\|\hat{A} - A\|^2 \leq \frac{Cp_1p_2(p_1\vee p_2)}{m_1m_2}$ with high probability if we treat $Y$ as the approximate rank-$r$ matrix.
		\item When there are no structural missingness, i.e. $m_1 = p_1, m_2 = p_2$, the proposed method and theoretical analysis lead to
	\[
	\E\|\hat{A} - A\|_F^2 \leq \frac{(p_1\vee p_2)r\log(p_1\vee p_2)}{\theta_0}.
	\]
	 Compared with the result in \cite{chatterjee2015matrix}, an additional logarithm terms appear in the upper bound since we impose a weaker sub-Gaussian norm bound (as described in Condition \ref{con:sub-gaussian}) comparing to their uniform boundedness condition ($|Y_{ij}|\leq M$ almost surely).
		\item When there are neither sporadic or structural missingness, we have
	\[
	\E\|\hat{A} - A\|_F^2 \leq Cr(p_1\vee p_2), 
	\]
	which is of the same rate as in \cite{donoho2014minimax,gavish2014optimal}.
	\end{itemize}
\end{remark}

\section{Proofs}
\label{sec.proofs}

We collect the technical proofs to the main results of the paper in this section.
\subsection{Proof of Theorem \ref{th:upper_bound}.} 

There are two options when implementing Algorithm \ref{al:procedure}, here we focus on the row thresholding situation as the other case be derived similarly by symmetry. We plan to prove the theorem by steps. 

\noindent (\emph{Step 1}) 
We define $U_{(1)} = U_{[\Omega_1, :]}$, $V_{(1)} = V_{[\Omega_2, :]}$. For the rest of the proof, we assume that the following inequalities hold,
\begin{equation}\label{ineq:assumption1_th1}
\sigma_{r}(U_{(1)}) \geq 3/4(m_1/p_1)^{1/2},\quad \sigma_r(U_{(2)}) \geq 3/4(m_2/p_2)^{1/2},
\end{equation}
\begin{equation}\label{ineq:assumption1_th2}
\left\|\tilde{Y}_{(11)} - A_{(11)}\right\| \leq \frac{9}{80}\left( \frac{C_{gap}(m_1+m_2)\log m}{\theta_0}\right)^{1/2},
\end{equation}
\begin{equation}\label{ineq:assumption1_th3}
\left\|\tilde{Y}_{(\bullet 1)} - A_{(\bullet 1)}\right\| \leq  \frac{1}{10}\left(\frac{C_{gap}(p_1+m_2)\log p}{\theta_0}\right)^{1/2},
\end{equation}
\begin{equation}\label{ineq:assumption1_th4}
\left\|\tilde{Y}_{(1 \bullet)} - A_{(1 \bullet)}\right\| \leq  \frac{1}{10}\left(\frac{C_{gap}(m_1+p_2)\log p}{\theta_0}\right)^{1/2}.
\end{equation}

\noindent (\emph{Step 2}) With Assumptions \eqref{ineq:assumption1_th1} - \eqref{ineq:assumption1_th4}, in this step we prove that $\hat{r} \geq r$, i.e. the stopping criterion is met when $s = r$. We only need to show that
\begin{equation}\label{eq:B_{21, r}B_{11, r}^-1}
\begin{split}
\left\|B_{21, [:, 1:r]} B_{11, [1:r, 1:r]}^{-1} \right\| \leq 2\sqrt{p_1/m_1}.
\end{split}
\end{equation}
First, we shall note that $A_{(11)} = U_{(1)}\Sigma V_{(1)}^\top, A_{(1\bullet)} = U_{(1)}\Sigma V^\top$, and $A_{(\bullet 1)}U\Sigma V_{(1)}^\top$, \eqref{ineq:assumption1_th1} -- \eqref{ineq:assumption1_th4} and Lemma 4 in \cite{zhang2016cross} imply that
\begin{equation}\label{ineq:sigma_r_A_11}
\begin{split}
\sigma_{r}(A_{(11)}) \geq & \sigma_r(U_{(1)}\Sigma_{(1)}) \sigma_r(U_{(1)}^\top) \geq \sigma_{\min}(U_{(1)})\sigma_{\min}(\Sigma)\sigma_{\min}(V_{(1)}) \\
\geq & \frac{9}{16}\cdot \left(\frac{m_1m_2}{p_1p_2}\right)^{1/2} \cdot \left(C_{gap}\log p\left(\frac{p_1p_2}{m_1\theta_0} + \frac{p_1p_2}{m_2\theta_0}\right)\right)^{1/2}\\
\geq & \frac{9}{16} \left(\frac{C_{gap}\log p(m_1 + m_2)}{\theta_0}\right)^{1/2}\\
\geq & \frac{1}{5} \left\|\tilde{Y}_{(11)} - A_{(11)}\right\|.
\end{split}
\end{equation}
Similarly one can also show that
\begin{equation}\label{ineq:sigma_r_A_bullet1}
\begin{split}
& \sigma_r(A_{(\bullet 1)}) \geq \frac{3}{4} \left(\frac{C_{gap}(p_1+m_2)\log p}{\theta_0}\right)^{1/2} \geq \frac{1}{5}\|\tilde{Y}_{(\bullet 1)} - A_{(\bullet 1)}\|,\\ 
& \sigma_r(A_{(1 \bullet)}) \geq \frac{3}{4} \left(\frac{C_{gap}(m_1+p_2)\log p}{\theta_0}\right)^{1/2} \geq \frac{1}{5}\|\tilde{Y}_{(1 \bullet)} - A_{(1 \bullet)}\|.
\end{split}
\end{equation}
Since $\rank(A_{(1\bullet)}) = \rank(A_{(\bullet1)}) = r$, we also have
\begin{equation*}
\begin{split}
& \sigma_{r+1}(\tilde{Y}_{(\bullet 1)}) = \min_{\rank(M) \leq r} \|\tilde{Y}_{(\bullet 1)} - M\| \leq \|\tilde{Y}_{(\bullet 1)} - A_{(\bullet 1)}\|,\\
& \sigma_{r+1}(\tilde{Y}_{(1 \bullet)}) = \min_{\rank(M) \leq r} \|\tilde{Y}_{(1 \bullet)} - M\| \leq \|\tilde{Y}_{(1 \bullet)} - A_{(1 \bullet)}\|.
\end{split}
\end{equation*}
Denote $K\in \mathbb{O}_{m_1, r}, L\in \mathbb{O}_{m_2, r}$ as the left and right singular vectors of $A_{(11)}$, respectively. Clearly, $K, L$ are also the left singular subspace of $A_{(1\bullet)}$ and the right singular subspace fo $A_{(\bullet 1)}$, respectively. Also, we denote $\hat{K} = U^{(1)}_{[:, 1:r]}$ and $\hat{L} = V^{(2)}_{[:, 1:r]}$ as the first $r$ left and right singular vectors of $\tilde{Y}_{(1\bullet)}$ and $\tilde{Y}_{(\bullet  1)}$, respectively. By the perturbation bound (Proposition 1 in \cite{cai2017rate}), we have the following upper bound for $K$ and the first $r$ columns of $\hat{K}$,
\begin{equation*}
\left(1 - \sigma_{\min}^2(\hat{K}^\top K)\right)^{1/2} = \left\|\sin\Theta(\hat{K}, K)\right\|\leq \frac{\sigma_{r}(A_{(1\bullet)})\|\tilde{Y}_{(1\bullet)} - A_{(1\bullet)}\|}{\sigma_r^2(A_{(1\bullet)}) - \sigma_{r+1}^2(\tilde{Y}_{(1\bullet)})} \leq \frac{1/5 \cdot 1/5}{1 - 1/5^2} = \frac{1}{24}, 
\end{equation*}
which means $\sigma_{\min}\left(\hat{K}^\top K\right) \geq \sqrt{575/576}$. Similarly we can also show that $\sigma_{\min}\left(\hat{L}^\top L\right) \geq \sqrt{575/576}$. 

Now we consider the following decomposition for $B_{11, [1:r, 1:r]}$ and $B_{\bullet 1, [:, 1:r]}$. According to the definition of $B_{11}$ in the proposed procedure,
\begin{equation}\label{eq:def_bar-A}
\begin{split}
B_{11, [1:r, 1:r]} = & \hat{K}^\top \tilde{Y}_{(11)} \hat{L} = \hat{K}^\top A_{(11)} \hat{L} + \hat{K}^\top \left(\tilde{Y}_{(11)} - A_{(11)}\right)\hat{L} :=  \bar{A}_{(11)} + \bar{Z}_{(11)},\\
B_{\bullet 1, [:, 1:r]} = & Y_{(\bullet 1)}\hat{L} = A_{(\bullet 1)} \hat{L} + (\tilde{Y}_{(\bullet 1)} - A_{(\bullet 1)})\hat{L} := \bar{A}_{(\bullet 1)} + \bar{Z}_{(\bullet 1)}.
\end{split}
\end{equation}
For $\bar{A}_{(11)}, \bar{Z}_{(11)}, \bar{A}_{(\bullet1)}$, and $\bar{Z}_{(\bullet1)}$, we have
\begin{equation*}
\begin{split}
\sigma_r\left(\bar{A}_{(11)}\right) = & \sigma_{r}\left(\hat{K}^\top A_{(11)}\hat{L}\right) = \sigma_{r}\left(\hat{K}^\top K K^\top A_{(11)} L L^\top \hat{L}\right)\\
\geq & \sigma_{\min}(\hat{K}^\top K) \sigma_r(K^\top A_{(11)}L)\sigma_{\min}(L^\top \hat{L})\geq \frac{575}{576}\sigma_r(A_{(11)}),
\end{split}
\end{equation*}
\begin{equation}\label{ineq:bar_Z11_bar_A11}
\|\bar{Z}_{11}\| \leq \|\tilde{Y}_{(11)} - A_{(11)}\| \leq \frac{1}{5} \sigma_r(A_{(11)}),
\end{equation}
\begin{equation}\label{ineq:bar_Z_upper_bound}
\|\bar{Z}_{(\bullet 1)}\| \leq \|\tilde{Y}_{(\bullet 1)} - Z_{(\bullet 1)}\| \overset{\eqref{ineq:assumption1_th3}}{\leq} \frac{1}{10}\left(\frac{C_{gap} (p_1 + m_2) \log p}{\theta_0}\right)^{1/2} \overset{\eqref{ineq:sigma_r_A_11}}{\leq}\frac{8}{45}\sqrt{\frac{p_1}{m_1}}\sigma_{r}(A_{(11)}).
\end{equation}
\begin{equation*}
\begin{split}
\bar{A}_{(\bullet 1)}\bar{A}_{(11)}^{-1} = & A_{(\bullet 1)} \hat{L}\left(\hat{K}^\top A_{(11)} \hat{L}\right)^{-1} = U\Sigma V_{(1)}^\top \hat{L} \left(\hat{K}^\top U_{(1)} \Sigma V_{(1)}^\top \hat{L}\right)^{-1}\\
= & U\Sigma V_{(1)}^\top \hat{L}\left(V_{(1)}^\top \hat{L}_{[:, 1:r]}\right)^{-1} \Sigma^{-1} \left(\hat{K}^\top U_{(1)}\right)^{-1}\\
= & U \left(\hat{K}^\top \mathbb{P}_K U_{(1)}\right)^{-1} \quad \text{since $U_{(1)}$  and $K$ share the same column space}\\
\leq & U(K^\top U_{(1)})^{-1} (\hat{K}^\top K)^{-1} 
\end{split}
\end{equation*}
Thus, 
\begin{equation}\label{ineq:bar_A_bar_A_11_inverse}
\|\bar{A}_{(\bullet 1)} \bar{A}_{(11)}^{-1}\| \leq \sigma_{\min}^{-1}(U_{(1)}) \sigma_{\min}^{-1} (\hat{K}^\top K) \leq 4/3\cdot \sqrt{575/576} \cdot (p_1/m_1)^{1/2}.
\end{equation}
Similarly we also 
\begin{equation}\label{ineq:bar_A1bullet_bar_A_11_inverse}
\|\bar{A}_{(11)}^{-1}\bar{A}_{(1\bullet)} \| \leq 4/3\cdot \sqrt{575/576} \cdot (p_2/m_2)^{1/2}.
\end{equation}
Now we are ready to verify \eqref{eq:B_{21, r}B_{11, r}^-1},
\begin{equation*}
\begin{split}
& \left\|B_{\bullet 1, [:, 1:r]} B_{11, [1:r, 1:r]}^{-1}\right\| = \left\|(\bar{A}_{(\bullet 1)} + \bar{Z}_{\bullet 1})(\bar{A}_{(11)} + \bar{Z}_{11})^{-1}\right\|\\
\leq & \left\|\bar{A}_{(\bullet 1)}(\bar{A}_{(11)} + \bar{Z}_{11})^{-1}\right\| + \left\|\bar{Z}_{\bullet 1}(\bar{A}_{(11)} + \bar{Z}_{11})^{-1}\right\|\\
\leq & \left\|\bar{A}_{(\bullet 1)}\bar{A}_{(11)}^{-1}\left(I - \bar{Z}_{(11)}(\bar{A}_{(11)} + \bar{Z}_{11})^{-1}\right)\right\| + \|\bar{Z}_{(\bullet 1)}\| \cdot \left(\sigma_{\min}(\bar{A}_{(11)}) - \|\bar{Z}_{11}\|\right)^{-1}\\
\overset{\eqref{ineq:bar_Z_upper_bound}\eqref{ineq:bar_Z11_bar_A11}}{\leq} & \left\|\bar{A}_{(\bullet 1)}\bar{A}_{(11)}^{-1}\right\| \cdot \left(1 + \|\bar{Z}_{(11)}\| \cdot \left(\sigma_{\min}(\bar{A}_{(11)}) - \|\bar{Z}_{11}\|\right)^{-1}\right)\\
& + \frac{4}{15}\sqrt{\frac{p_1}{m_1}}\sigma_{r}(A_{(11)}) \cdot \sigma_{r}^{-1}(A_{(11)})\left(\frac{575}{576}-\frac{1}{5}\right)^{-1}  \\
\overset{\eqref{ineq:bar_A_bar_A_11_inverse}}{\leq} & \frac{4}{3}\left(\frac{575}{576}\right)^{1/2} \sqrt{\frac{p_1}{m_1}} \left(1 +\frac{1}{5} \left(\frac{575}{576} - \frac{1}{5}\right)^{-1}\right) + \frac{8}{45}\sqrt{\frac{p_1}{m_1}}\left(\frac{575}{576}-\frac{1}{5}\right)^{-1}\\
\leq & 2 \sqrt{\frac{p_1}{m_1}}.
\end{split}
\end{equation*}
Note that $B_{21, [:, 1:r]} B_{11, [1:r, 1:r]}^{-1}$ is a sub-matrix of $B_{\bullet 1, [:, 1:r]} B_{11, [1:r, 1:r]}^{-1}$, so that $\|B_{21, [:, 1:r]} B_{11, [1:r, 1:r]}^{-1}\| \leq \|B_{\bullet 1, [:, 1:r]} B_{11, [1:r, 1:r]}^{-1}\|$, and we have finished the proof for \eqref{eq:B_{21, r}B_{11, r}^-1}. Based on the definition of the iterative algorithm, we must have $\hat{r} \geq r$.

\ \par

\noindent (\emph{Step 3}) In order to analyze the upper bound for $\|\hat{A} - A\|$, we introduce the following notations
\begin{equation*}
\begin{split}
& B_L = B_{21, [:, 1:\hat{r}]} = \tilde{Y}_{(\bullet 1)} V^{(2)\top}_{[1:\hat{r}, :]} \in \mathbb{R}^{(p_1-m_1)\times \hat{r}}, \quad B_R = B_{12, [:, 1:\hat{r}]} =  U^{(1)}_{[1:\hat{r}, :]}\tilde{Y}_{1\bullet}\in \mathbb{R}^{\hat{r}\times (p_2-m_2)}\\
& B_M = B_{11, [1:\hat{r}, 1:\hat{r}]} = U^{(1)}_{[1:\hat{r}, :]}\tilde{Y}_{11}V^{(2)\top}_{[1:\hat{r}, :]} \in \mathbb{R}^{\hat{r}\times \hat{r}};
\end{split}
\end{equation*}
\begin{equation*}
\begin{split}
& A_L = A_{(21)} V^{(2)\top}_{[1:r, :]} \in \mathbb{R}^{(p_1-m_1)\times r}, \quad A_R =   U^{(1)}_{[1:r, :]}A_{(12), [1:r, 1:r]}\in \mathbb{R}^{r\times p_2}\\
& A_M = U^{(1)}_{[1:r, :]}A_{(11), [1:r, 1:r]}V^{(2)\top}_{[1:r, :]} \in \mathbb{R}^{r\times r}.
\end{split}
\end{equation*}
	\begin{equation*}
	\tilde{A}_L = 
	\begin{bmatrix}
	A_L & 0_{p_1\times (\hat{r} - r)}
	\end{bmatrix},\quad \tilde{A}_M = \begin{bmatrix}
	A_M & 0_{r \times (\hat{r} - r)}\\
	0_{(\hat{r} - r)\times r} & 0_{(\hat{r} - r)\times(\hat{r} - r)}
	\end{bmatrix}, \quad \tilde{A}_R = \begin{bmatrix}
	A_R\\
	0_{(\hat{r} - r)\times p_2}
	\end{bmatrix}
	\end{equation*}
With the assumption \eqref{ineq:assumption1_th1} -- \eqref{ineq:assumption1_th4} and the intermediate conclusion that $\hat{r} \geq r$, in this step, we plan to prove the following inequalities and then apply Lemma \ref{lm:A_LA_MA_R-Y_LY_MY_R} to develop the upper bound of \eqref{ineq:upper_bound}: 
\begin{equation}\label{ineq:to-be-prove-1}
\|\tilde{A}_L - B_L\| \leq C\left(\log p \left(\frac{p_1+m_2}{\theta_0}\right)\right)^{1/2}, \quad \|\tilde{A}_R - B_R\| \leq C\left(\log p \left(\frac{p_2+m_1}{\theta_0}\right)\right)^{1/2},
\end{equation}
\begin{equation}\label{ineq:to-be-prove-2}
\|\tilde{A}_M - B_M\| \leq C\left(\log m \left(\frac{m_1+m_2}{\theta_0}\right)\right)^{1/2}, \quad \|\tilde{A}_M - B_M\| \leq \frac{1}{4}\sigma_{\min}(A_M).
\end{equation}
\begin{equation}\label{ineq:to-be-prove-3}
\|A_LA_M^{-1}\| \leq 2\sqrt{\frac{p_1}{m_1}}, \quad \|B_LB_M^{-1}\| \leq 2\sqrt{\frac{p_1}{m_1}},\quad \|A_M^{-1}A_R\| \leq 2\sqrt{\frac{p_2}{m_2}}.
\end{equation}
For the first part of \eqref{ineq:to-be-prove-1}, we have
\begin{equation*}
\begin{split}
& \left\|\tilde{A}_L  - B_L\right\| = \left\|\begin{bmatrix}
\left(\tilde{Y}_{(21)} - A_{(21)}\right)\hat{L} & \tilde{Y}_{(21)} V^{(2)}_{[(r+1):\hat{r}, :]}
\end{bmatrix}
\right\|\\
\leq & \left(\|\tilde{Y}_{(21)} - A_{(21)}\|^2 + \|\tilde{Y}_{(\bullet 1)} V^{(2)}_{[(r+1):\hat{r}, :]}\|^2 \right)^{1/2}\leq \left(\|\tilde{Y}_{(21)} - A_{(21)}\|^2 + \sigma_{r+1}^2(\tilde{Y}_{(\bullet 1)}) \right)^{1/2}\\
\leq & \left(\|\tilde{Y}_{(\bullet 1)} - A_{(\bullet 1)}\|^2 + \|\tilde{Y}_{(\bullet 1)} V^{(2)}_{[(r+1):\hat{r}, :]}\|^2 \right)^{1/2}\leq \left(\|\tilde{Y}_{(21)} - A_{(21)}\|^2 + \|\tilde{Y}_{(21)} - A_{(21)}\|^2 \right)^{1/2}\\
\overset{\eqref{ineq:assumption1_th3}}{\leq} & C\left(\frac{(p_1+m_2)\log p}{\theta_0}\right)^{1/2}
\end{split}
\end{equation*}
The second part of \eqref{ineq:to-be-prove-1} and \eqref{ineq:to-be-prove-2} can be proved similarly. Next, $A_L, A_M$ are exactly $\bar{A}_{(21)}$ and $\bar{A}_{(11)}$ defined as \eqref{eq:def_bar-A}, thus $\|A_L A_M^{-1}\| \leq 2\sqrt{p_1/m_1}$. Similarly, $\|A_M^{-1} A_R\| \leq 2\sqrt{p_2/m_2}$. By the statement of assumption, we also know that $\|B_LB_M^{-1}\| = \|B_{21, [:, 1:\hat{r}]}B_{11, [:, 1:\hat{r}]}^{-1}\|\leq 2\sqrt{p_1/m_1}$. Therefore we have proved \eqref{ineq:to-be-prove-3}.

\ \par

\noindent (\emph{Step 4}) Combining \eqref{ineq:to-be-prove-1}, \eqref{ineq:to-be-prove-2}, and \eqref{ineq:to-be-prove-3}, Lemma \ref{lm:A_LA_MA_R-Y_LY_MY_R} implies 
$$\|B_LB_M^{-1} B_R - A_L A_M^{-1} A_R\| \leq C\left(\frac{\log p}{\theta_0}\left(\frac{p_1p_2}{m_1}+\frac{p_1p_2}{m_2}\right)\right)^{1/2}.$$ 
Note that $B_LB_M^{-1} B_R = \hat{A}_{(22)}$, also $A_L A_M^{-1} A_R = A_{(22)}$ by Theorem 1 in \cite{cai2016structured}. we have proved the following upper bound for $\hat{A}_{(22)} - A_{(22)}$,
$$\|\hat{A}_{(22)} - A_{(22)}\| \leq C\left(\frac{\log p}{\theta_0}\left(\frac{p_1p_2}{m_1}+\frac{p_1p_2}{m_2}\right)\right)^{1/2}.$$ 
Now it remains to consider the error bound $\hat{A}_{(21)} - A_{(21)}$, $\hat{A}_{(12)} - A_{(12)}$, and $\hat{A}_{(11)} - A_{(11)}$. In fact,
\begin{equation*}
\begin{split}
& \|\hat{A}_{(21)} - A_{(21)}\| = \left\|\tilde{Y}_{(21)} V^{(2)}_{[:, 1:\hat{r}]}V^{(2)\top}_{[:, 1:\hat{r}]} - A_{(21)}\right\|\\
= & \left\|\tilde{Y}_{(21)} - A_{(21)} + \tilde{Y}_{(21)} V^{(2)}_{[:, \hat{r}+1:m_2]}V^{(2)\top}_{[:, \hat{r}+1:m_2]} \right\|\\
\leq & \|\tilde{Y}_{(21)} - A_{(21)}\| + \|\sigma_{r+1}(\tilde{Y}_{(\bullet 1)}) \|\leq 2 \|\tilde{Y}_{(21)} - A_{(21)}\| \leq C\left(\frac{\log p}{\theta} (p_1+m_2)\right).
\end{split}
\end{equation*}
We can similarly derive the upper bound for $\hat{A}_{(12)} - A_{(12)}$ and $\hat{A}_{(11)} - A_{(11)}$. To sum up, under the assumptions \eqref{ineq:assumption1_th1}--\eqref{ineq:assumption1_th4}, one can prove that
$$\|\hat{A} - A\| \leq C\left(\frac{\log p}{\theta_0}\left(\frac{p_1p_2}{m_1}+\frac{p_1p_2}{m_2}\right)\right)^{1/2}.$$ 
for some uniform constant $C>0$.

\ \par

\noindent (\emph{Step 5}) In this final step, we analyze the probability that \eqref{ineq:assumption1_th1} - \eqref{ineq:assumption1_th4} hold. By Lemma \ref{lm:tilde_Y_perturbation}, \eqref{ineq:assumption1_th2} - \eqref{ineq:assumption1_th4} hold with probability at least $1 - m_1^\alpha - m_2^{\alpha}$ provided large enough $C_{gap}$. By Lemma \ref{lm:U_Omega_upper_lower_bound}, \eqref{ineq:assumption1_th1} holds with probability $1 - 2r\exp(-m_1/(64\rho r)) - 2r\exp(-m_2/(64\rho r))$. By Condition 1, i.e. sub-Gaussian observations, $|A_{ij}| = |EY_{ij}| \leq E|Y_{ij}| \leq \tau$, thus $\|A\|_F^2 \leq p_1p_2\tau^2$. On the other hand, since $\rank(A) = r$, $\|A\|_F^2 \geq r\sigma_{\min}^2(A) = C_{gap}\left(\frac{rp_1p_2}{m_1} + \frac{rp_1p_2}{m_2}\right)\log p$, thus we have $C_{gap} \log p\left(\frac{r}{m_1} + \frac{r}{m_2}\right) \leq 1,$ which implies $\min\{m_1/r, m_2/r\} \geq C_{gap} \log p$. Note that $p \geq r$,thus for large enough constant $C_{gap}$, we have
$$1 - 2r\exp(-m_1/(64\rho r)) - 2r\exp(-m_2/(64\rho r)) \geq 1 - 4r \exp\left(\frac{C_{gap}\log p}{64\rho}\right) \geq 1 - Cp^{-\alpha}.$$
In summary, 
\begin{equation}
P\left(\text{\eqref{ineq:assumption1_th1} - \eqref{ineq:assumption1_th4} all hold} \right) \geq 1 - Cp^{-\alpha}
\end{equation}
for some constant $C, \alpha>0$. \quad $\square$

\subsection{Proof of Theorem \ref{th:lower_bound}.} 

For better presentation, we prove this theorem by steps.

\begin{enumerate}[leftmargin=*]

\item In this step, we construct a series of $A^{(k)}\in \mathbb{R}^{p_1\times p_2}$. Without loss of generality, assume $\Omega_1 = \{1:m_1\}$, $\Omega_2 = \{1:m_2\}$. Let $r = 1$ and generate $A_0 = \frac{1}{2} 1_{p_1} 1_{p_2}^\top$, and randomly generate $\mu^{(k)} \in \mathbb{R}^{p_1}$ such that
$$\mu^{(k)}_{[1:m_1]} = 1, \quad \mu^{(k)}_{[(m_1+1): p_1]} \overset{iid}{\sim} \left\{\begin{array}{ll}
1 , & \text{with probability } \frac{1}{2};\\
-1, &  \text{with probability } \frac{1}{2}.
\end{array}\right. $$ 
Then we construct
\begin{equation*}
A^{(k)} = A_0 + \lambda\mu^{(k)} 1_{p_2}^\top, \quad k = 1,\ldots, N,
\end{equation*}
where $\lambda>0$ is some constant to be determined later. Based on this construction, we can see for $k\neq l$,
\begin{equation*}
\mu^{(k)} - \mu^{(l)} = \sum_{i=m_1+1}^{p_1} \left(\mu^{(k)}_i - \mu^{(l)}_i\right)^2, \quad \left(\mu^{(k)}_i - \mu^{(l)}_i\right)\overset{iid}{\sim} \left\{\begin{array}{ll}
4, \quad \text{w. p. } \frac{1}{2},\\
0, \quad \text{w. p. } \frac{1}{2},\\
\end{array}\right. \quad m_1+1 \leq i \leq p_1.
\end{equation*}
By Bernstein's inequality,
\begin{equation}
\begin{split}
& P\left(\left\|\mu^{(k)} - \mu^{(l)}\right\|_2^2 \geq 3(p_1-m_1) \text{ or } \left\|\mu^{(k)} - \mu^{(l)}\right\|_2^2 \leq (p_1-m_1)\right)\\
= & P\left(\left|\sum_{i=1}^{p_1} \left\{\left(\mu^{(k)}_i - \mu^{(l)}_i\right)^2 - 2\right\}\right| \geq (p_1-m_1) \right)\\
\leq & 2\exp\left(- \frac{(p_1-m_1)^2/2}{4(p_1-m_1) + \frac{2}{3}p_2}\right) = 2\exp\left(-3(p_1-m_1)/28\right).
\end{split}
\end{equation}
Therefore, whenever $N < \frac{1}{2}\exp(3(p_1-m_1)/28)$, there is a positive probability that 
\begin{equation}\label{ineq:condition_mu^k_mu^l}
p_1-m_1\leq \min_{1\leq k < l \leq N}\left\{\|\mu^{(k)} - \mu^{(l)}\|_2^2\right\} \leq \max_{1\leq k < l \leq N}\left\{\|\mu^{(k)} - \mu^{(l)}\|_2^2\right\} \leq  3(p_1-m_1),
\end{equation}
which means there exist a set of fixed $\{\mu^{(k)}\}_{k=1}^{N}$ such that \eqref{ineq:condition_mu^k_mu^l} holds. For the rest of the proof, we suppose $\{\mu^{(1)},\ldots, \mu^{(N)}\}$ are such choice, and $N = \lceil \frac{1}{2}\exp\left(3(p_1 - m_1)/28\right)\rceil - 1$. 

\item Next, we construct the observations $Y^{(k)}\odot M^{(k)}$. We specifically set all entries of $\Theta_{(11)}$ and $\Theta_{(12)}$ to be 1, all entries of $\Theta_{(22)} = 0$, and all entries of $\Theta_{(21)}$ to be $\theta_0$. Then we construct the missingness indicators as $M_{ij} \sim \text{Bernoulli}(\Theta_{ij})$ independently, where $1\leq i \leq p_1, 1\leq j \leq p_2$, and suppose $Y^{(k)} = A^{(k)} + Z^{(k)}$, while $Z^{(k)} \overset{iid}{\sim} N(0, \frac{1}{4})$. $\{M^{(k)}\}_{k=1}^N \sim \text{Bernoulli}(\Theta)$. Our finally observations are $Y^{(k)}\odot M^{(k)}$. Moreover, we can check that $A^{(k)}$, $M^{(k)}$ and $Y^{(k)}\odot M^{(k)}$ satisfy Conditions 1, 2, 3, and 4.

\item Now, we focus on the distribution of $Y^{(k)}\odot M^{(k)}$: $\mathcal{P}_{Y^{(k)}\odot M^{(k)}}$, where we aim to show $\mathcal{P}_{Y^{(k)}}$ are close to each other while $A^{(k)}$'s are far apart. The central technique relies on generalized Fano's lemma (see, e.g., Theorem 1 in \cite{yang1999information}).

It is well known that the Kullback-Leibler divergence between two Gaussian vectors is 
$$KL\left(N\left(\mu_A, \sigma^2 I\right), N\left(\mu_B, \sigma^2 I\right)\right) = \frac{1}{2\sigma^2}\left\|\mu_A - \mu_B\right\|_2^2. $$
Then one can have the following calculation for the KL divergence of $\mathcal{P}_{\mathcal{P}_{Y^{(k)}\odot M^{(k)}}}$ and $\mathcal{P}_{Y^{(l)} \odot M^{(l)}}$ for any $1\leq k< l \leq N$,
\begin{equation*}
\begin{split}
& KL\left(\mathcal{P}_{Y^{(k)}\odot M^{(k)}}, \mathcal{P}_{Y^{(l)} \odot M^{(l)}}\right) = \sum_{R \in \{0, 1\}^{p_1\times p_2}} P(M = R) \cdot KL(\mathcal{P}_{Y^{(k)}\odot R}, \mathcal{P}_{Y^{(l)}\odot R})\\
= & \sum_{R\in \{0, 1\}^{p_1\times p_2}} P(M = R) \cdot \left\|\left(A^{(k)} - A^{(l)}\right)\odot R\right\|_F^2 \\
= & \sum_{R\in \{0, 1\}^{p_1\times p_2}} P(M = R) \cdot \sum_{i=1}^{p_1}\sum_{j=1}^{p_2}\left(A^{(k)}_{ij} - A^{(l)}_{ij}\right)^2\cdot R_{ij}\\
= & \sum_{i=m_1+1}^{p_1}\sum_{j=1}^{p_2} \left(A^{(k)}_{ij} - A^{(l)}_{ij}\right)^2 \E R_{ij} = \theta_0\left\|A^{(k)}_{(21)} - A^{(l)}_{(21)}\right\|_F^2\\
= & \theta_0 \lambda^2 m_2 \|\mu_k - \mu_l\|_2^2 \leq 3\theta_0 m_2(p_1-m_1) \lambda^2.
\end{split}
\end{equation*}
Here we used the fact that $A_{ij}^{(k)}$ and $A_{ij}^{(l)}$ are different only when $m_2+1\leq j \leq p_2$.

\item Finally, we apply the generalized Fano's lemma to establish the lower bound result. For any $1\leq k < l \leq N$,
\begin{equation}
\begin{split}
\left\|A^{(k)} - A^{(l)}\right\|^2 = \left\|\lambda(\mu^{(k)} - \mu^{(l)}) 1_{p_2}^\top \right\|^2 = \lambda^2 p_2 \|\mu^{(k)} - \mu^{(l)}\|_2 \geq \lambda^2 (p_1-m_1)p_2.
\end{split}
\end{equation}
By generalized Fano's lemma, 
\begin{equation}\label{ineq:generalized-Fano-intermediate}
\inf_{\hat{A}}\max_{1\leq k \leq N} \E \|\hat{A} - A^{(k)} \| \geq \frac{\lambda^2 (p_1-m_1)p_2}{2}\left(1 - \frac{3\theta_0m_2(p_1-m_1)\lambda^2 + \log 2}{\log N}\right)
\end{equation}
Now we set $\lambda^2 = \frac{\frac{1}{2}\log N - \log 2}{3\theta_0m_2(p_1-m_1)}$. Whenever $m_1 \geq 25$, $p_1-m_1 \geq 2m_1 - m_2 \geq 25$, then
$$\lambda^2 = \frac{\frac{1}{2}\log\left(\lceil \frac{1}{2}\exp(3(p_1 - m_1)/28) \rceil -1\right) - \log 2}{3\theta_0m_2(p_1-m_1)} \geq \frac{(p_1-m_1)/10}{3\theta_0m_2(p_1-m_1)} = \frac{1}{30\theta_0m_2} $$
\eqref{ineq:generalized-Fano-intermediate} yields
\begin{equation}
\inf_{\hat{A}} \sup_{\substack{\text{Conditions}\\ \text{1, 2, 3, 4 are net}}} \E \|\hat{A} - A^{(k)} \| \geq \frac{\lambda^2 (p_1-m_1)p_2}{2}\cdot \frac{1}{2} \geq \frac{p_1p_2}{120m_2}.
\end{equation}
By symmetry, we can similarly show
$$
\inf_{\hat{A}} \sup_{\substack{\text{Conditions}\\ \text{1, 2, 3, 4 are net}}} \E \|\hat{A} - A^{(k)} \| \geq \frac{p_1p_2}{120 m_1}.
$$
Therefore, we have finished the proof of Theorem \ref{th:lower_bound}. \quad $\square$
\end{enumerate}

\section{Technical Lemmas}

We collect the technical lemmas to the proofs of this paper in this section. 

\begin{Lemma}\label{lm:rank-one-matrix-expression}
	Suppose $A\in \mathbb{R}^{m\times p}$, $\rank(A) = 1$. If $\sum_{i,j} A_{ij} \neq0$, then 
	$$A_{ij} = \left(\sum_{i'=1}^m A_{i'j}\right)\left(\sum_{j'=1}^p A_{ij'}\right)/\left(\sum_{i'j'}A_{i'j'}\right), \quad 1\leq i \leq m, 1\leq j \leq p.$$
\end{Lemma}

{\bf\noindent Proof of Lemma \ref{lm:rank-one-matrix-expression}.} When $A$ is of rank-1, there exist two vectors $u\in\mathbb{R}^m, v\in \mathbb{R}^p$ such that $A = uv^\top$. Since
$$\sum_{i=1}^m\sum_{j=1}^p A_{ij} = \left(\sum_{i=1}^m u_i\right)\left(\sum_{j=1}^n v_j\right) \neq 0, $$
we have $\sum u_i \neq 0$ or $\sum v_j \neq 0$. Therefore,
\begin{equation*}
\begin{split}
 \left(\sum_{i'=1}^m A_{i'j}\right)\left(\sum_{j'=1}^p A_{ij'}\right)/\left(\sum_{i'j'}A_{i'j'}\right) = \frac{u_i (\sum_{j'} v_{j'})\cdot v_j(\sum_{i'}u_{i'})}{(\sum_i u_i)\cdot(\sum_j v_j)} = u_iv_j = A_{ij},
\end{split}
\end{equation*}
for all $1\leq i \leq m, 1\leq j \leq p$, which finished the proof for this lemma. \quad $\square$

\

\begin{Lemma}\label{lm:tilde_Y_perturbation}
	Recall the definition of $\hat{\Theta}, \tilde{Y}$ in . Given the assumptions in the proof for Theorem \ref{th:upper_bound}, we have the following perturbation bounds for $\tilde{Y}_{(11)}$, $\tilde{Y}_{(\bullet 1)}$, and $\tilde{Y}_{(1\bullet)}$ with probability at least $1 - p^{-\alpha}$:
	\begin{equation*}
	\left\|\tilde{Y}_{(11)} - A_{(11)}\right\| \leq C\sqrt{\frac{(m_1+m_2)\log p}{\theta_0}},
	\end{equation*}
	\begin{equation*}
	\left\|\tilde{Y}_{(1\bullet)} - A_{(1\bullet)}\right\| \leq C\sqrt{\frac{(m_1+p_2)\log p}{\theta_0}},\quad \left\|\tilde{Y}_{(\bullet 1)} - A_{(\bullet 1)}\right\| \leq C\sqrt{\frac{(m_2+p_1)\log p}{\theta_0}}.
	\end{equation*}
	Furthermore, if all entries of $A_{(11)}, A_{(12)}$, and $A_{(21)}$ are observable without missing (but with possibly noise), we have the following more accurate upper bounds:
	\begin{equation*}
	\left\|\tilde{Y}_{(11)} - A_{(11)}\right\| \leq C\sqrt{\frac{m_1+m_2}{\theta_0}},
	\end{equation*}
	\begin{equation*}
	\left\|\tilde{Y}_{(1\bullet)} - A_{(1\bullet)}\right\| \leq C\sqrt{\frac{m_1+p_2}{\theta_0}},\quad \left\|\tilde{Y}_{(\bullet 1)} - A_{(\bullet 1)}\right\| \leq C\sqrt{\frac{m_2+p_1}{\theta_0}}.
	\end{equation*}
	with probability at least $1 - m_1^{-\alpha} - m_2^{-\alpha}$.
\end{Lemma}

{\noindent\bf Proof of Lemma \ref{lm:tilde_Y_perturbation}.} 
For convenience, we denote $m = m_1+m_2$, $p = p_1+p_2$. First we introduce the following notations to bridge $\tilde{Y}$ and $A$:
\begin{equation}\label{eq:def_tilde_Y^0}
\begin{split}
& \tilde{Y}_{ij}^0 := Y_{ij}\oslash \theta_{ij},\quad 1\leq i \leq p_1, 1\leq j\leq p_2.\\
& \tilde{Y}_{(11)}^0 = (\tilde{Y}_{ij}^0)_{i\in \Omega_1, j\in \Omega_2}, \quad \tilde{Y}_{(12)}^0 = (\tilde{Y}_{ij}^0)_{i\in \Omega_1, j\in \Omega_2^c},\quad \tilde{Y}_{(21)}^0 = (\tilde{Y}_{ij}^0)_{i\in \Omega_1^c, j\in \Omega_2}.
\end{split}
\end{equation}
Then we clearly have $\left\|\tilde{Y}_{(kl)} - A_{(kl)}\right\| \leq \left\|\tilde{Y}_{(kl)} - \tilde{Y}^0_{(kl)}\right\| + \left\|\tilde{Y}^0_{(kl)} - A_{(kl)}\right\|,$ for all $1\leq k, l\leq 2$. To develop the error bounds for $\tilde{Y}_{(11)}$, $\tilde{Y}_{(12)}$, and $\tilde{Y}_{(21)}$, it is sufficient to prove the following inequalities.
\begin{equation}\label{ineq:Y_11^0-A_11}
\begin{split}
& \left\|\tilde{Y}_{(11)}^0 - A_{(11)}\right\| \leq C\sqrt{\frac{(m_1+m_2)\log p}{\theta_0}}, \\
\left\|\tilde{Y}_{(\bullet 1)}^0 - A_{(\bullet 1)}\right\|&  \leq C\sqrt{\frac{(p_1+m_2)\log p}{\theta_0}},\quad \left\|\tilde{Y}_{(1\bullet)}^0 - A_{(1\bullet)} \right\| \leq C\sqrt{\frac{(m_1+p_2)\log p}{\theta_0}}.
\end{split}
\end{equation}
\begin{equation}\label{ineq:Y_11^0-Y_11}
\begin{split}
& \left\|\tilde{Y}_{(11)}^0 - \tilde{Y}_{(11)}\right\| \leq \sqrt{\frac{(m_1+m_2)\log p}{\theta_0}},\\
\left\|\tilde{Y}_{(\bullet 1)}^0 - \tilde{Y}_{(\bullet 1)}\right\|&  \leq \sqrt{\frac{(p_1+m_2)\log p}{\theta_0}},\quad \left\|\tilde{Y}_{(1\bullet)}^0 - \tilde{Y}_{(1\bullet)} \right\| \leq \sqrt{\frac{(m_1+p_2)\log p}{\theta_0}}.
\end{split}
\end{equation}	
with probability at least $1 - O\left(p^{-\alpha}\right)$. 

\begin{enumerate}[leftmargin=*]
\item In this step, we aim to show \eqref{ineq:Y_11^0-A_11} with probability at least $1 - O(p^{-\alpha})$. We first introduce the following lemma on matrix concentration bounds provided by \cite{koltchinskii2011nuclear}.
\begin{Proposition}[\cite{koltchinskii2011nuclear}, Proposition 2]\label{pr:matrix-concentration}
	Let $A_1,\ldots, A_n$ be independent random matrices with dimensions $p_1$-by-$p_2$ satisfying $E(A_i) = 0$ and $\|A_i\|\leq U$ almost surely for $i=1,\ldots, n$. Here $U$ is some uniform constant. Define
	$$\sigma_A = \max\left\{\left\|\frac{1}{n}\sum_{k=1}^n \E(A_kA_k^\top)\right\|^{1/2}, \left\|\frac{1}{n}\sum_{k=1}^n \E(A_k^\top A_k)\right\|^{1/2}\right\}. $$
	$$U_A^{(\alpha)} = \inf\{u>0: \E \exp(\|A\|^{\alpha}/u^\alpha) \leq 2\}, \quad \alpha \geq 1. $$
	Then for all $t>0$, with probability at least $1 - e^{-t}$ we have
	\begin{equation*}
	\left\|\frac{1}{n}\sum_{k=1}^nA_k\right\| \leq C\max\left\{\sigma_A \sqrt{\frac{t+\log(m)}{n}}, U_Z^{(\alpha)}\left(\log \frac{U_A^{(\alpha)}}{\sigma_A}\right)\frac{t+\log m}{n}\right\},
	\end{equation*}
	where $m = m_1+m_2$.
\end{Proposition}
Now let us consider the proof for \eqref{ineq:Y_11^0-A_11}. Denote $e_i^{(m)}$ as the $i$-th canonical basis in $m$-dimensional space, i.e. $e_i^{(m)}$ is a $m$-dimensional vector with $i$-th entry as 1 and others as 0. Then
\begin{equation*}
\tilde{Y}_{(11)} - A_{(11)} = \sum_{i=1}^{m_1} \sum_{j=1}^{m_2} \left(\tilde{Y}_{ij} - A_{ij}\right)e_i^{(m_1)} e_j^{(m_2)\top} := \sum_{i=1}^{m_1}\sum_{j=1}^{m_2} E_{ij},
\end{equation*}
where $E^{(ij)} = \left(\tilde{Y}_{ij} - A_{ij}\right)e_i^{(m_1)} e_j^{(m_2)\top} \in \mathbb{R}^{m_1\times m_2}$. For any two symmetric matrices $A, B$, we write $A \preceq B$ if and only if $B - A $ is positive semi-definite. Then
\begin{equation*}
\begin{split}
& \sum_{i=1}^{m_1}\sum_{j=1}^{m_2} \E E^{(ij)} E^{(ij)\top} = \sum_{i=1}^{m_1}\sum_{j=1}^{m_2}e_i^{(m_1)}e_i^{(m_1)\top}\E(\tilde{Y}_{ij} - A_{ij})^2\\
= & \sum_{i=1}^{m_1} e_i^{(m_1)}e_i^{(m_1)\top} \sum_{j=1}^{m_2} \left(\theta_{ij}\left(A_{ij}^2 + \sigma_{ij}^2\right) -\theta_{ij}^2A_{ij}^2 \right)/\theta_{ij}^2\\
\preceq & \sum_{i=1}{m_1} \sum_{j=1}^{m_2} e_i^{(m_1)}e_i^{(m_1)\top} \E(A_{ij} + Z_{ij})^2 / \theta_{ij} \preceq \sum_{i=1}^{m_1} \sum_{j=1}^{m_2} e_i^{(m_1)}e_i^{(m_1)\top} \E(A_{ij} + Z_{ij})^2 / \theta_{0} \preceq \frac{m_2}{\theta_0}I_{m_1}.
\end{split}
\end{equation*}
Similarly one can also show that $\sum_{i=1}^{m_1}\sum_{j=1}^{m_2} E^{(ij)\top} E^{(ij)} \preceq \frac{m_1}{\theta_0}I_{m_2}$. Moreover, 
$$U^{(2)} = \inf\left\{u>0: \E\exp(\|E\|^2/u^2) \leq 2\right\} \leq C;$$
$$\sigma_A = \left\|\frac{1}{m_1m_2} \sum_{i=1}^{m_1}\sum_{j=1}^{m_2} \E E^{(ij)} E^{(ij)\top}\right\| \geq \frac{c}{m_1m_2}.$$
By Proposition \ref{pr:matrix-concentration}, where we set $n = m_1m_2$, $t = C\log(p)$, and $Z_1,\ldots, Z_n = E_{11},\ldots, E_{m_1,m_2}$ we have with probability at least $1 - p^{-\alpha}$,
\begin{equation}
\begin{split}
\left\|\frac{1}{m_1m_2}\sum_{i=1}^{m_1}\sum_{j=1}^{m_2} \left(\tilde{Y}_{(11)}^{0} - A_{(11)} \right)\right\| \leq & C\max\Big\{\left(\frac{1}{m_1\theta_0} + \frac{1}{m_2\theta_0}\right)^{1/2} \cdot \sqrt{\frac{\log(p)}{m_1m_2}}, \\
& \log\left(\frac{m_1m_2}{\theta_0}\right)\cdot \frac{\log(m_1+m_2)}{m_1m_2} \Big\}.
\end{split}
\end{equation}
Therefore we have
\begin{equation}
\left\|\tilde{Y}_{(11)}^0 - A_{(11)}\right\|\leq  C\left(\frac{(m_1+m_2)\log p}{\theta_0}\right)^{1/2}.
\end{equation}
with probability at least $1 - p^{-\alpha}$. We can similarly also write down the proof for the upper bound of $\|\tilde{Y}_{(\bullet 1)}^0 - A_{(\bullet 1)}\|$ and $\|\tilde{Y}_{(1\bullet)} ^0 - A_{(1\bullet)}\|$. Thus, we have finished the proof for \eqref{ineq:Y_11^0-A_11}.

\item We target to show \eqref{ineq:Y_11^0-Y_11} in this step. First let us focus on developing the upper bound for $\|\tilde{Y}_{(11)}^0 - \tilde{Y}_{(11)}\|$. By definition of $\tilde{Y}$ and $\tilde{Y}^0$ \eqref{eq:def_tilde_Y^0}, we have
\begin{equation}\label{eq:tildeY^0-tildeY}
\begin{split}
\left\|\tilde{Y}_{(11)}^0 - \tilde{Y}_{(11)}\right\| = \left\|Y_{(11)} \odot \left(1 \oslash \hat{\Theta}_{(11)} - 1\oslash \Theta_{(11)}\right)\right\| = \left\|\tilde{Y}_{(11)}\odot \left(\Theta_{(11)}\oslash \hat{\Theta}_{(11)}- 1\right)\right\|.
\end{split}
\end{equation}
For convenience, we introduce the following notation that represents the number of observed entries in each row and column and their expectation correspondingly,
\begin{equation}\label{eq:def_M_Theta_bullet}
\begin{split}
& M_{i\bullet}^{(1)} = \sum_{j=1}^{p_2}M_{ij},\quad M_{\bullet j}^{(1)} = \sum_{i=1}^{m_1}M_{ij},\quad M_{\bullet \bullet}^{(1)} = \sum_{i=1}^{m_1}\sum_{j=1}^{p_2}M_{ij},\\
& \Theta_{i\bullet}^{(1)} = \sum_{j=1}^{p_2}M_{ij},\quad \Theta_{\bullet j}^{(1)} = \sum_{i=1}^{m_1}M_{ij},\quad \Theta_{\bullet \bullet}^{(1)} = \sum_{i=1}^{m_1}\sum_{j=1}^{p_2}M_{ij}.
\end{split}
\end{equation}
Particularly due to $\rank(\Theta)  =1$, it is easy to show the following fact,
\begin{equation}
\theta_{ij} = \frac{\Theta_{\bullet j}^{(1)}\Theta_{i\bullet}^{(1)}}{\Theta_{\bullet\bullet}^{(1)}}, \quad 1\leq i \leq m_1, 1\leq j \leq p_2.
\end{equation}
Since $M_{ij} \sim {\rm Bernoulli}(\theta_{ij})$, we have $\E M_{ij} = \theta_{ij}, \Var(M_{ij}) = \theta_{ij}(1-\theta_{ij})$, and $|M_{ij}-\theta_{ij}| \leq 1$. Then Bernstein's inequality yields
\begin{equation}\label{ineq:M_bulletj_bernstein}
P\left(M_{\bullet j}^{(1)} - \Theta^{(1)}_{\bullet j}\right) = P\left(\left|\sum_{i=1}^{m_1} M_{ij} - \sum_{i=1}^{m_1} \theta_{ij}\right| \geq t \right) \leq 2\exp\left(- \frac{t^2/2}{\sum_{i=1}^{m_1}\theta_{ij}(1-\theta_{ij}) + t/3}\right).
\end{equation}
Note that $\theta_{ij} \geq \frac{C_{gap}r\log p}{m_1\wedge m_2}$, we know 
\begin{equation}\label{ineq:Theta_ibullet_lower_bound}
\Theta_{i\bullet}^{(1)} = \sum_{j=1}^{p_2}\theta_{ij} \geq p_2\theta_0\geq C_{gap}r\log p, \quad  \Theta_{\bullet j}^{(1)} = \sum_{i=1}^{m_1}\theta_{ij} \geq m_1\theta_0 \geq C_{gap}r\log p.
\end{equation}
We specifically let $t = C\sqrt{\Theta_{i\bullet}^{(1)}\log p}$ for some large constant $C>0$ in \eqref{ineq:M_bulletj_bernstein}, then
\begin{equation*}
\begin{split}
& P\left(|M_{i\bullet}^{(1)} - \Theta_{i\bullet}^{(1)} |\leq C\sqrt{\Theta_{i\bullet}^{(1)} \log p} \right) \geq 1 - 2\exp\left(-\frac{C^2 \Theta_{i\bullet}\log p}{\Theta_{i\bullet}^{(1)}+C\sqrt{\Theta_{i\bullet}^{(1)}\log p}/3}\right)\geq 1 - p^{-\alpha}.
\end{split}
\end{equation*}
Meanwhile, we shall also note from \eqref{ineq:Theta_ibullet_lower_bound} that
\begin{equation}\label{ineq:M_bulleti-Theta}
|M_{i\bullet}^{(1)} - \Theta_{i\bullet}^{(1)} |\leq C\sqrt{\Theta_{i\bullet}^{(1)} \log p} \quad \Rightarrow \quad |M_{i\bullet}^{(1)} - \Theta_{i\bullet}^{(1)}| \leq \frac{1}{3}\Theta_{i\bullet}^{(1)}
\end{equation}
when $C_{gap}$ is a large enough constant. Similarly one can also show with probability at least $1 - p^{-\alpha}$,
\begin{equation}\label{ineq:M_bulletj-Theta}
|M_{\bullet j}^{(1)} - \Theta_{\bullet j}^{(1)} |\leq C\sqrt{\Theta_{\bullet j}^{(1)} \log p} \leq \frac{1}{3} \Theta_{\bullet j}^{(1)}, \quad |M_{\bullet\bullet}^{(1)} - \Theta_{\bullet\bullet}^{(1)} |\leq C\sqrt{\Theta_{\bullet\bullet}^{(1)} \log p} \leq \frac{1}{3}\Theta_{\bullet\bullet}^{(1)}. 
\end{equation}
Thus, with probability at least $1 - 3p^{-\alpha}$, we have \eqref{ineq:M_bulleti-Theta} and \eqref{ineq:M_bulletj-Theta} both hold, then for any $i, j$,
\begin{equation*}
\begin{split}
& \left|1 - \theta_{ij}/\hat{\Theta}_{ij}^{(1)}\right| = \left|1 - \frac{M_{\bullet\bullet}^{(1)}\Theta_{\bullet j}^{(1)}\Theta_{i \bullet}^{(1)}}{M_{\bullet j}^{(1)} M_{i\bullet}^{(1)}\Theta_{\bullet\bullet}^{(1)}}\right| = \left| \frac{\Theta_{\bullet\bullet}^{(1)}M_{\bullet j}^{(1)}M_{i \bullet}^{(1)} - M_{\bullet\bullet}^{(1)}\Theta_{\bullet j}^{(1)}\Theta_{i \bullet}^{(1)}}{M_{\bullet j}^{(1)}M_{i\bullet}^{(1)}\Theta_{\bullet\bullet}^{(1)}}\right|\\
\leq & \frac{\left|\left(M_{\bullet\bullet}^{(1)} - \Theta_{\bullet\bullet}^{(1)} \right)\Theta_{\bullet j}^{(1)}\Theta_{i\bullet}^{(1)}\right| +  \left|\Theta_{\bullet\bullet}^{(1)} \left(M_{\bullet j}^{(1)} - \Theta_{\bullet j}^{(1)}\right)\cdot M_{i\bullet}^{(1)}\right| + \left|\Theta_{\bullet\bullet}^{(1)}M_{\bullet j}^{(1)}(M_{i\bullet}^{(1)} - \Theta_{i\bullet}^{(1)})\right|}{M_{\bullet j}^{(1)} M_{i\bullet}^{(1)} \Theta_{\bullet\bullet}^{(1)}}\\
\overset{\eqref{ineq:M_bulleti-Theta}\eqref{ineq:M_bulletj-Theta}}{\leq} & C\frac{\sqrt{\log p \Theta_{\bullet\bullet}^{(1)}} \Theta_{\bullet j}^{(1)}\Theta_{i\bullet}^{(1)} + \sqrt{\log p \Theta_{\bullet j}^{(1)}} \Theta_{\bullet\bullet}^{(1)}\Theta_{i\bullet}^{(1)} + \sqrt{\log p \Theta_{i \bullet}^{(1)}} \Theta_{\bullet j}^{(1)}\Theta_{\bullet\bullet}^{(1)}}{\Theta_{\bullet j}^{(1)} \Theta_{i\bullet}^{(1)} \Theta_{\bullet\bullet}^{(1)}}\\
\overset{\eqref{ineq:Theta_ibullet_lower_bound}}{\leq} & C\sqrt{\frac{\log p}{\theta_0}\left(\frac{1}{m_1}+\frac{1}{p_2}\right)}.
\end{split}
\end{equation*}
Similarly one can also show $|1-\theta_{ij}/\hat{\Theta}_{ij}^{(2)}| \leq C\sqrt{\frac{\log p}{\theta_0}(m_2^{-1}\wedge p_1^{-1})}$, $\forall i, j$ with probability at least $1 - p^{-\alpha}$. By Step 1 of the proof of this lemma, we also have $\|\tilde{Y}^0_{(11)} - A_{(11)}\| \leq C\sqrt{\frac{(m_1+m_2)\log p}{\theta_0}}$ with probability at least $1 - p^{-\alpha}$. Also, $\|A_{(11)}\| \leq \sqrt{\sum_{i=1}^{m_1}\sum_{j=1}^{m_2}A_{ij}^2} \leq C\sqrt{m_1m_2}$.
Therefore, with probability at least $1 - p^{-\alpha}$,
\begin{equation*}
\begin{split}
& \left\|\tilde{Y}^0_{(11)} - \tilde{Y}_{(11)}\right\| \overset{\eqref{eq:tildeY^0-tildeY}}{\leq} \|\tilde{Y}^0_{(11)}\|\cdot C\sqrt{\frac{\log p}{\theta_0}\left(\frac{1}{m_1} + \frac{1}{m_2}\right)}\\
\leq & C\sqrt{\frac{\log p}{\theta_0}\left(\frac{1}{m_1} + \frac{1}{m_2}\right)} \left(\|A_{(11)}\| + \|\tilde{Y}^0_{(11)} - A_{(11)}\|\right)\\
\leq & C\sqrt{\frac{\log p}{\theta_0}\left(\frac{1}{m_1} + \frac{1}{m_2}\right)} \left(C\sqrt{m_1m_2} + C\sqrt{\frac{\log p (m_1+m_2)}{\theta_0}}\right) \\
\leq & C\sqrt{\frac{(m_1+m_2)\log p}{\theta_0}} \quad (\text{since } \theta_0\geq \frac{C_{gap}r\log p}{m_1\wedge m_2}).
\end{split}
\end{equation*}
Next, one can similarly show that
\begin{equation*}
\left\|\tilde{Y}^0_{(\bullet 1)} - \tilde{Y}_{(\bullet 1)}\right\| \leq C\sqrt{\frac{(p_1+m_2)\log p}{\theta_0}};\quad \left\|\tilde{Y}^0_{(1\bullet)} - \tilde{Y}_{(1\bullet)}\right\| \leq C\sqrt{\frac{(m_1+p_2)\log p}{\theta_0}}.
\end{equation*}
with probability at least $1 - p^{-\alpha}$, which has finished the proof for \eqref{ineq:Y_11^0-Y_11}.
\end{enumerate}

To conclude from Steps 1 and 2, we have proved the error bound for $\tilde{Y}_{(11)}$, $\tilde{Y}_{(1\bullet)}$, and $\tilde{Y}_{(\bullet 1)}$. 
\begin{equation*}
\left\|\tilde{Y}_{(11)} - A_{(11)}\right\| \leq C\sqrt{\frac{(m_1+m_2)\log p}{\theta_0}},
\end{equation*}
\begin{equation*}
\left\|\tilde{Y}_{(1\bullet)} - A_{(1\bullet)}\right\| \leq C\sqrt{\frac{(m_1+p_2)\log p}{\theta_0}},\quad \left\|\tilde{Y}_{(\bullet 1)} - A_{(\bullet 1)}\right\| \leq C\sqrt{\frac{(m_2+p_1)\log p}{\theta_0}}.
\end{equation*}
with probability at least $1 - p^{-\alpha}$. \quad $\square$

The following lemma, as a special case of Lemma 2 in \cite{zhang2016cross}, provides the upper and lower bound for submatrix of orthogonal matrices.

\

\begin{Lemma}\label{lm:U_Omega_upper_lower_bound}
	Suppose $U\in \mathbb{O}_{p, r}$ is an orthogonal matrix, and $U$ satisfies the incoherence condition with constant $\rho$, i.e. $\frac{r}{p} \|\mathbb{P}_Ue_i\| \leq \rho$, $\forall 1\leq i \leq p$. Suppose we uniformly randomly select $m$ numbers from $\{1, \ldots, p\}$ and form them as the set $\Omega$, then for any $m > r$, $0 < \varepsilon < 1$,
	\begin{equation}
	P\left(\frac{(1-\varepsilon)m}{p}\leq \sigma_{\min}(U_{[\Omega, :]}) \leq \sigma_{\max}(U_{[\Omega, :]}) \leq \frac{\varepsilon^2 m}{p} \right) \geq 1 - 2r\exp(-m\varepsilon^2/(4\rho r)).
	\end{equation}	
\end{Lemma}

The next Lemma \ref{lm:A_LA_MA_R-Y_LY_MY_R} provides the upper bound for $A_LA_M^{-1} A_R - B_L B_M^{-1} B_R$ when the difference of each pair of $\{A_L, B_L\}$, $\{A_M, B_M\}$, and $\{A_R, B_R\}$ are bounded.

\

\begin{Lemma}\label{lm:A_LA_MA_R-Y_LY_MY_R}
	Suppose $\hat{r} \geq r$, $A_L\in \mathbb{R}^{p_1\times r}, A_M \in \mathbb{R}^{r\times r}, A_R\in \mathbb{R}^{r\times p_2},$; $B_L\in \mathbb{R}^{p_1\times \hat{r}}, B_M \in \mathbb{R}^{\hat{r}\times\hat{r}}, B_R\in \mathbb{R}^{\hat{r}\times \hat{r}}$. We further denote
	\begin{equation*}
	\tilde{A}_L = 
	\begin{bmatrix}
	A_L & 0_{p_1\times (\hat{r} - r)}
	\end{bmatrix},\quad \tilde{A}_M = \begin{bmatrix}
	A_M & 0_{r \times (\hat{r} - r)}\\
	0_{(\hat{r} - r)\times r} & 0_{(\hat{r} - r)\times(\hat{r} - r)}
	\end{bmatrix}, \quad \tilde{A}_R = \begin{bmatrix}
	A_R\\
	0_{(\hat{r} - r)\times p_2}
	\end{bmatrix}
	\end{equation*}
	and assume that $\|\tilde{A}_L - B_L\| \leq \alpha_L$, $\|\tilde{A}_M - B_M\| \leq \alpha_M$, and $\|\tilde{A}_R - B_R\| \leq \alpha_R$.
	Provided that $\|Z_M\| \leq \frac{1}{4}\sigma_{\min}(A_M)$, $\|Z_L\| \leq C\lambda_L\sigma_{\min}(A_M)$, $\|Z_R\| \leq C\lambda_R \sigma_{\min}(A_M)$, and
	\begin{equation}
	\|A_L A_M^{-1}\| \leq \lambda_L, \quad \|B_L B_M^{-1}\| \leq \lambda_L, \quad \|A_M^{-1} A_R\|\leq \lambda_R,
	\end{equation} 
	(note that there is assumption on $\|B_M^{-1}B_R\|$), for some constant $C>0$ we have
	\begin{equation*}
	\left\|A_L A_M^{-1} A_{R} - B_L B_M^{-1} B_R\right\| \leq C\left(\lambda_L\alpha_R + \lambda_R\alpha_L + \lambda_L\lambda_R\alpha_M\right).
	\end{equation*}
\end{Lemma}

{\bf\noindent Proof of Lemma \ref{lm:A_LA_MA_R-Y_LY_MY_R}.} The lemma can be proved in steps. For convenience, we denote $Z_L = B_L - A_L$, $Z_M = B_M - A_M$, and $Z_R = B_R - A_R$.
\begin{enumerate}[leftmargin=*]
	\item We first assume that the SVD of $B_M$ is $B_M = U\Sigma V^\top$. Here  $$U = \begin{bmatrix}
	U_1 & U_2
	\end{bmatrix} = \begin{bmatrix}
	U_{11} & U_{12}\\
	U_{21} & U_{22}
	\end{bmatrix}, \quad \Sigma = \begin{bmatrix}
	\Sigma_1 & \\
	& \Sigma_2
	\end{bmatrix}, \quad V = \begin{bmatrix}
	V_1 & V_2
	\end{bmatrix} = \begin{bmatrix}
	V_{11} & V_{12}\\
	V_{21} & V_{22}
	\end{bmatrix}.$$
	Denote 
	$$W = \begin{bmatrix}
	I_r\\
	0_{(\hat{r} - r)\times r}
	\end{bmatrix} \in \mathbb{O}_{\hat{r}, r},$$
	then $W$ is clearly the left and right singular subspaces of $\tilde{A}_M$. By the perturbation bound result in Proposition 1 of \cite{cai2017rate}, we have
	\begin{equation*}
	\begin{split}
	& \left\|\sin\Theta(U_1, W)\right\| \leq \frac{\sigma_{r+1}(B_M)\|Z_M\|}{\sigma_{\min}^2(B_M W) - \|\sigma_{r+1}(B_M)}\\
	\leq & \frac{\|Z_M\|^2}{(\sigma_{\min}(A_M) - \|Z_M\|)^2 - \|Z_M\|^2} \leq \frac{1}{3^2 - 1} = \frac{1}{8}.
	\end{split}
	\end{equation*}
	Here we used the condition that $\sigma_{\min}(A_M) > 4\|Z_M\|$. We can similarly prove that $\|\sin\Theta(V_1, W)\| \leq 1/8$. Based on the properties of $\sin\Theta$ distance (Lemma 1 in \cite{cai2016structured}), we have $\|U_{12}\| = \|U^\top W_{\perp}\| \leq 1/8$, 
	\begin{equation}
	\sigma_{\min}\left(U_{11}\right) = \sqrt{1 - \|U^\top W_{\perp}\|^2} \geq \frac{\sqrt{63}}{8}.
	\end{equation}
	\item We denote $\mathring{B}_L = B_LV_1$, $\mathring{B}_M = U_1^\top B_MV_1$, $\mathring{B}_R = U_1^\top B_R$. We can define $\mathring{A}_L, \mathring{A}_M, \mathring{A}_R, \mathring{Z}_L, \mathring{Z}_L$, and $\mathring{Z}_L$ in the same fashion. Then one can calculate that
	\begin{equation}
	\|\mathring{Z}_L\| \leq \alpha_L, \quad \|\mathring{Z}_M\| \leq \alpha_M, \quad \|\mathring{Z}_R\| \leq \alpha_R.
	\end{equation}
	\begin{equation}\label{ineq:sigma_min_A_B}
	\begin{split}
	& \sigma_{\min}\left(\mathring{A}_M\right) \geq \sigma_{\min}(U_{11})\sigma_{\min}(A_M)\sigma_{\min}(V_{11}) \geq \frac{63}{64}\sigma_{\min}(A_M) \geq \frac{63}{16}\|Z_M\|, \\
	& \sigma_{\min}\left(\mathring{B}_M\right) \geq \sigma_{\min}\left(\mathring{A}_M\right) - \|Z_M\| \geq \frac{47}{63}\sigma_{\min}(\mathring{A}_M) \geq \frac{47}{16}\|Z_M\|.
	\end{split}
	\end{equation}
	Based on that the SVD of $B_M$ is written as $B_M = U\Sigma V^\top$, we have the following forms for $B_LB_M^{-1}B_R$ and $A_LA_M^{-1}A_R$, which will be useful in our later analysis.
	\begin{equation}\label{eq:lm_decomposition_B}
	\begin{split}
	& B_LB_M^{-1}B_R = B_LV\Sigma^{-1} U^\top B_R = B_L(V_1 \Sigma_1^{-1} U_1^\top + V_2\Sigma_2^{-1} U_2^\top) B_R\\
	= & B_LV_1\Sigma_1^{-1}U_1^\top B_R + B_LV_2\Sigma_2^{-1}U_2^\top B_R\\
	= & B_LV_1\left(U_1^\top B_M V_1\right)^{-1} U_1^\top B_R + B_L V_2\left(U_2^\top B_M V_2\right)^{-1} U_2^\top B_R\\
	= & \mathring{B}_L\mathring{B}_M^{-1} \mathring{B}_R + B_L V_2\left(U_2^\top B_M V_2\right)^{-1} U_2^\top B_R.
	\end{split}
	\end{equation}
	Similarly, $A_LA_M^{-1}A_R$ can be rewritten as
	\begin{equation}\label{eq:lm_decomposition_A}
	\begin{split}
	& A_LA_M^{-1}A_R = A_LV_{11} \left(U_{11}^\top A_M V_{11}\right)^{-1} U_{11} A_R = \tilde{A}_LV_{1} \left(U_{1}^\top \tilde{A}_M V_{1}\right)^{-1} U_{1} \tilde{A}_R = \mathring{A}_L \mathring{A}_M^{-1} \mathring{A}_R.
	\end{split}
	\end{equation}
	\item In this step, we develop the following properties for $\mathring{B}_L, \mathring{B}_M, \mathring{B}_R$ and $\mathring{A}_L, \mathring{A}_M, \mathring{A}_R$:
	\begin{equation}\label{ineq:lemma-to-be-prove1}
	\|\mathring{B}_L\mathring{B}_M^{-1} \| \leq C\lambda_L, \quad \|\mathring{B}_M^{-1}\mathring{B}_R^{-1} \| \leq C\lambda_R,
	\end{equation}
	\begin{equation}\label{ineq:lemma-to-be-prove2}
	\|B_L V_2\left(U_2B_MV_2\right)^{-1} U_2^\top B_R\| \leq C\lambda_L\alpha_R + C\lambda_L\lambda_R \alpha_M,
	\end{equation}
	\begin{equation}\label{ineq:lemma-to-be-prove3}
	\|\mathring{B}_L\mathring{B}_M^{-1} \mathring{B}_R - \mathring{A}_L\mathring{A}_M^{-1} \mathring{A}_R \| \leq C\lambda_R\alpha_L + C\lambda_R\alpha_R + C\lambda_L\lambda_R \alpha_M.
	\end{equation}
	First, \eqref{ineq:lemma-to-be-prove1} can be proved by
	\begin{equation}
	\begin{split}
	& \left\|\mathring{B}_L \mathring{B}_M^{-1}\right\| = \left\|\left(\mathring{A}_L+\mathring{Z}_L\right) \left(\mathring{A}_M + \mathring{Z}_M\right)^{-1}\right\|\\
	\leq & \|\mathring{A}_L (\mathring{A}_M + \mathring{Z}_M)^{-1} \| + \|\mathring{Z}_L (\mathring{A}_M + \mathring{Z}_M)^{-1} \|\\
	\leq & \|\mathring{A}_L\mathring{A}_M^{-1} (I - \mathring{Z}_M(\mathring{A}_M+\mathring{Z}_M))\| + \|\mathring{Z}_L\| \cdot \|\mathring{B}_M^{-1}\|\\
	\overset{\eqref{ineq:sigma_min_A_B}}{\leq} & \lambda_L \left(1 + \|\mathring{Z}_M\| \cdot \frac{16}{47}\|Z_M\|^{-1}\|\right) + C\lambda_L \lambda_{\min}(A_M) \cdot \frac{64}{63} \sigma_{\min}^{-1}(A_M)\\
	\leq & C\lambda_L.
	\end{split}
	\end{equation}
	One can similarly prove that $\left\|\mathring{B}_M^{-1}\mathring{B}_R\right\| \leq C\lambda_R$, thus we have \eqref{ineq:lemma-to-be-prove1}. Next we consider the proof for \eqref{ineq:lemma-to-be-prove2}. Since
	\begin{equation}
	\begin{split}
	& \|U_2^\top B_R\| \leq \|U_2^\top Z_R\| + \|U_2^\top \tilde{A}_R\| \leq \alpha_R + \|U^\top_{2, [1:r, :]} A_R\|\\
	\leq & \alpha_R + \left\|(U_{2, [1:r, :]}^{\top} A_M) (U_{2, [1:r, :]}^\top A_M)^{-1} U^\top_{2, [1:r, :]} A_R\right\|\\
	\leq & \alpha_R + \|U_{2, [1:r, :]}^\top A_M\| \cdot \|A_M^{-1} A_R\| \leq \alpha_R + \|U_2^\top \tilde{A}_M\| \lambda_R\\
	\leq & \alpha_R + \left(\|U_2^\top (A_M + Z_M) \| + \|U_2^\top Z_M\|\right) \cdot \lambda_R\\
	\leq & \alpha_R + \lambda_R\left(\sigma_{r+1}(A_M + Z_M) + \alpha_M\right)\\
	\leq & \alpha_R + \lambda_R\left(\|Z_M\| + \alpha_M\right) = \alpha_R + 2\lambda_R \alpha_M.
	\end{split}
	\end{equation}
	\begin{equation}
	\begin{split}
	&\left\|B_LV_2 \left(U_2 B_M V_2\right)^{-1} U_2^\top\right\| = \left\|B_L V \left(U^\top  B_M V\right)^{-1} U - B_L V_1 \left(U_1^\top  B_M V_1\right)^{-1} U_1\right\|\\
	\leq & \|B_LB_M^{-1}\| + \|\mathring{B}_L \mathring{B}_M^{-1}\| \leq C\lambda_L.
	\end{split}
	\end{equation}
	The two inequalities above yield \eqref{ineq:lemma-to-be-prove2}:
	\begin{equation*}
	\|B_L V_2\left(U_2B_MV_2\right)^{-1} U_2^\top B_R\| \leq \|B_L V_2\left(U_2B_MV_2\right)^{-1} U_2^\top\| \cdot\|U_2^\top B_R\| \leq C\lambda_L\alpha_R + C\lambda_L\lambda_R\alpha_M.
	\end{equation*}
	Finally, \eqref{ineq:lemma-to-be-prove3} can be proved by
	\begin{equation}
	\begin{split}
	& \|\mathring{B}_L\mathring{B}_M^{-1} \mathring{B}_R - \mathring{A}_L\mathring{A}_M^{-1} \mathring{A}_R \| \leq \|\mathring{B}_L\mathring{B}_M^{-1}\mathring{Z}_R\| + \|\mathring{B}_L\mathring{B}_M^{-1}\mathring{A}_R - \mathring{A}_L\mathring{A}_M^{-1}\mathring{A}_R\|\\
	\leq & C\lambda_L \|\mathring{Z}_R\| + \|\mathring{Z}_L \mathring{B}_M^{-1} \mathring{A}_R\| + \|\mathring{A}_L \left(\mathring{B}_M^{-1} - \mathring{A}_M^{-1}\right) \mathring{A}_R\|\\
	\leq & C\lambda_L\alpha_R + C\lambda_R\alpha_L + \|\mathring{A}_L \left(\mathring{B}_M^{-1} - \mathring{A}_M^{-1}\right) \mathring{A}_R \|\\
	\leq & C\lambda_R\alpha_R + C\lambda_R\alpha_L + \|\mathring{A}_L \mathring{A}_M \left(\mathring{Z}_M - \mathring{Z}_M \mathring{B}_M^{-1} \mathring{Z}_M\right)\mathring{A}_M \mathring{A}_R\|\\
	\leq & C\lambda_R\alpha_L + C\lambda_L\alpha_R + \lambda_L\lambda_R\left(\|\mathring{Z}_M\| + \|\mathring{Z}_M\|^2 \|\mathring{B}_M^{-1}\|\right)\\
	\leq & C\lambda_R\alpha_L + C\lambda_L\alpha_R + C\lambda_L\lambda_R\alpha_M.
	\end{split}
	\end{equation}
	\item Now we are ready to analyze $\|B_LB_M^{-1}B_R - A_LA_M^{-1}A_R\|$ based on the previous calculations:
	\begin{equation*}
	\begin{split}
	&\|B_LB_M^{-1}B_R - A_LA_M^{-1}A_R\|\\ 
	\overset{\eqref{eq:lm_decomposition_A}\eqref{eq:lm_decomposition_B}}{\leq} & \left\|\mathring{B}_L\mathring{B}_M^{-1} \mathring{B}_R - \mathring{A}_L\mathring{A}_M^{-1} \mathring{A}_R\right\| + \left\|B_L V_2\left(U_2^\top B_M V_2\right)^{-1} U_2^\top B_R\right\|\\
	\overset{\eqref{ineq:lemma-to-be-prove2}\eqref{ineq:lemma-to-be-prove3}}{\leq} & C \lambda_L \alpha_R + C\lambda_R \alpha_L + C\lambda_L \lambda_R \alpha_M,
	\end{split}
	\end{equation*}
	which has finished the proof of Lemma \ref{lm:A_LA_MA_R-Y_LY_MY_R}. \quad $\square$
\end{enumerate}

\section{Upper Bound of Ranks}\label{sec: upp_rank}
In Step 4, when there exist prior knowledge or estimations on the noise level $\tau$ and missing parameter $\theta_0$ (which will be defined later in Conditions \ref{con:sub-gaussian} and \ref{con:missingness}), say $\tilde{\tau}$ and $\tilde{\theta}_0$, we can apply hard singular value thresholding (HSVD) (see, e.g. 	\cite{donoho2014minimax,gavish2014optimal,chatterjee2015matrix}) to $B_{(\bullet 1)}$ and $B_{(1\bullet)}$ before locating $\hat{r}$:
	$$B_{(\bullet 1)} \Leftarrow HSVT_{\lambda_1} \left(B_{(\bullet 1)}\right), \quad B_{(1\bullet)} \Leftarrow HSVT_{\lambda_2}\left(B_{(1\bullet)}\right).$$
	Here, $\lambda_1 = \eta\sqrt{\frac{p_1\tilde{\tau}\log(p_1\vee p_2)}{\tilde{\theta}_0}}$, $\lambda_2 = \eta\sqrt{\frac{p_2\tilde{\tau}\log(p_1\vee p_2)}{\tilde{\theta}_0}}$, $\tilde{\tau}$ and $\tilde{\theta}_0$, $\eta>0$ is a constant; for any matrix with SVD: $D = \sum_i \sigma_i u_i v_i^\top$, $HSVD_{\lambda}(D):= \sum_i \sigma_i 1_{\{\sigma_i \geq \lambda\}}u_iv_i^\top$ is defined as the hard singular value thresholding operator. 
	Interestingly, this additional HSVT here is equivalent to starting with a smaller value $r_0$ for locating $\hat{r}$ in Step 4. To be specific, a reasonable choice of $r_0$ is
		\begin{equation}\label{eq:selection of r_0}
		r_0 = \min\left\{i: \sigma_i(\tilde{Y}_{(\bullet 1)}) \geq \eta\sqrt{\frac{p_1\tilde{\tau}\log(p_1\vee p_2)}{\tilde{\theta}_0}}, \text{ and } \sigma_i(\tilde{Y}_{(1\bullet)}) \geq \eta\sqrt{\frac{p_2\tilde{\tau}\log(p_1\vee p_2)}{\tilde{\theta}_0}}\right\}. 
		\end{equation}
		Furthermore, we also have the following ad hoc choices for $\tilde{\theta}_0$ and $\tilde{\tau}$,
		\begin{equation}\label{eq:theta_0_tau}
		\hat{\theta}_0 = \min\left\{\hat{\Theta}_{ij}, 1\leq i\leq m_1, \text{ or } 1\leq j \leq m_2\right\}; \quad \tilde{\tau} = \left(\sum_{i,j : M_{ij}=1} Y_{ij}^2\right)^{1/2} / \left(\sum_{i,j} M_{ij} \right)^{1/2}.
		\end{equation}
		We will further analyze such choice of $r_0$ in Section \ref{sec.theory}. On the other hand, when $\tau$ and $\theta_0$ are unavailable or hard to estimate, the regular choice $r_0 = m_1\wedge m_2$ without additional HSVT step will produce low-rank and good enough estimates in practice. 

\section{Supporting Results}

\subsection{Additional Simulations}\label{asup_eval}
As the low-rank assumption is essential for \textsc{Macomss} in denoising and imputation, we further examine the case where $A$ is only approximately low-rank to evaluate the robustness of this condition to our method. In the following setting, we fix $p_1 = p_2 = 300$, $r = 3$, and generate $A = U V^\top$, where $U \in \mathbb{R}^{p_1\times r}$ and $V \in \mathbb{R}^{p_2\times r}$ are uniformly random orthogonal matrices. 
\begin{enumerate}
    \item[1.] Assume that $\Theta$ and $Z$ are generated in the same way as in our first setting. Alternatively, we construct $A = U D V^\top$, where $U \in \mathbb{O}_{p_1, (p_1 \wedge p_2)}$ and $V \in \mathbb{O}_{p_2, (p_1 \wedge p_2)}$ are uniformly randomly generated orthogonal matrices, and $D = \mathrm{diag}(\underbrace{1, \ldots, 1}_r, 1^{-\alpha}, \ldots, (p_1 \wedge p_2))$. Clearly, $\alpha$ indicates the rate of singular value decay for $A$, as larger $\alpha$ values correspond to $A$ being closer to an exact low-rank matrix. 
\end{enumerate}

In addition to continuous-valued datasets, we consider a special setting where the entries of the matrix are discrete count observations. Count data matrices are prevalent in applications, such as EHR data analysis \citep{beaulieu2018characterizing,hemingway2018big}, microbiome studies \citep{shi2016regression,cao2016count}, astronomical energy spectra \citep{nowak2000statistical}, and fluorescence microscopy \citep{jiang2015minimax}, where the Poisson distribution is commonly used for modeling. We consider the following setting:
\begin{enumerate}
    \item[2.] We let $m_1 = m_2 \in \{10, 20, 50, 100\}$. The sporadic missing parameter $\Theta$ is generated in the same way as the first simulation setting. 
The observations are generated as:
$Y_{ij} \sim \text{Poisson}\left(\lambda \cdot A_{ij}\right)$,
where $A = UV^\top$, with $U \in \mathbb{R}^{p_1 \times r}$ and $V \in \mathbb{R}^{p_2 \times r}$. Each entry of $U$ and $V$ is independently sampled from $|N(0, 1/p_1)|$ and $|N(0, 1/p_2)|$, respectively. 
The intensity parameter $\lambda$ is set as: $\lambda = \lambda_0 \cdot \frac{p_1p_2}{\sum_{i,j} A_{ij}}$,
ensuring that the average value of each entry of $Y$ is approximately $\lambda_0$. We let $\lambda_0$ vary from 1 to 10, representing different signal-to-noise ratios (SNRs).
\end{enumerate}

\begin{figure}[h]
    \begin{center}
		\includegraphics[height = 2in,width=3in]{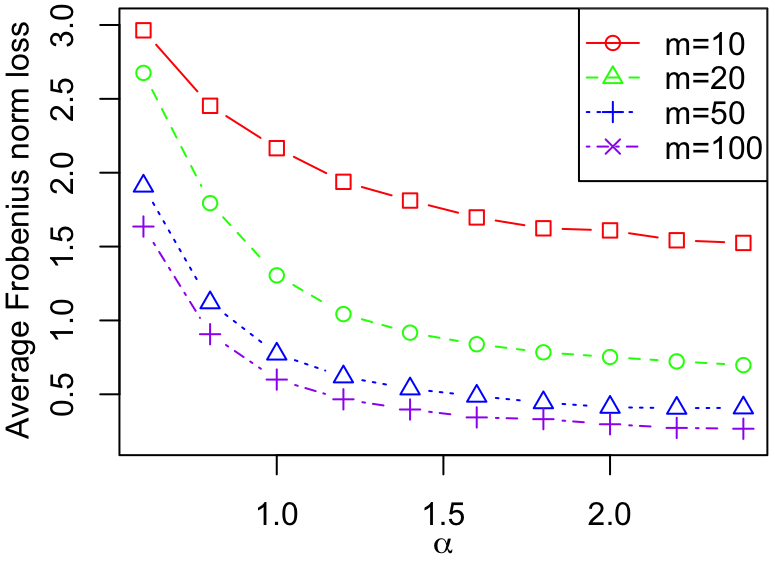} \includegraphics[height = 2in,width=3in]{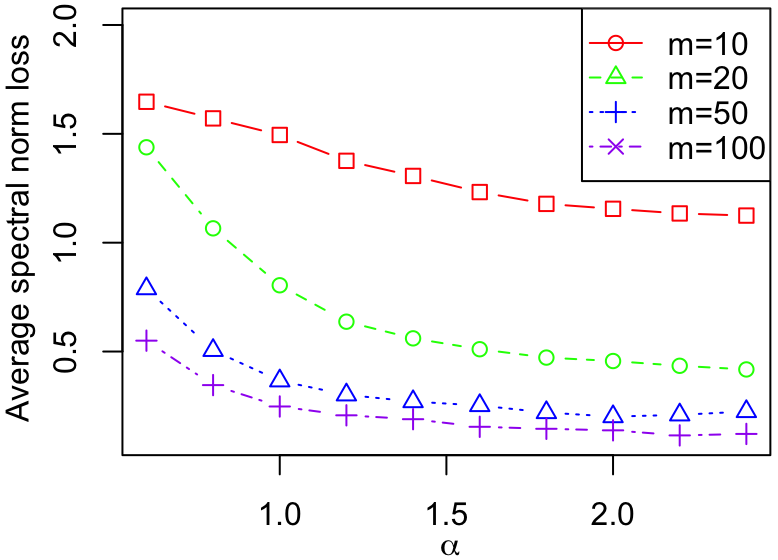}
		\caption{Average Frobenius and spectral loss for approximately low-rank matrix $A$.}
		\label{fig:simu-setting3}
	\end{center}
    \begin{center}
		\includegraphics[height = 2in,width=3in]{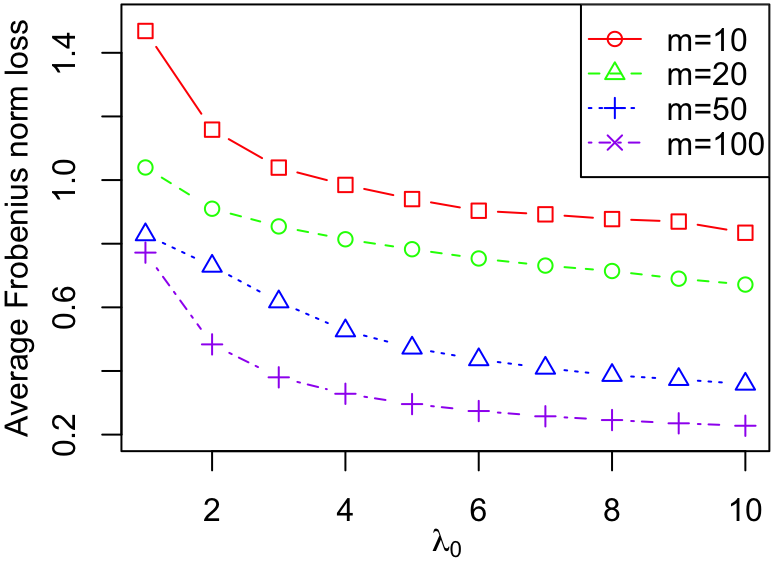} \includegraphics[height = 2in,width=3in]{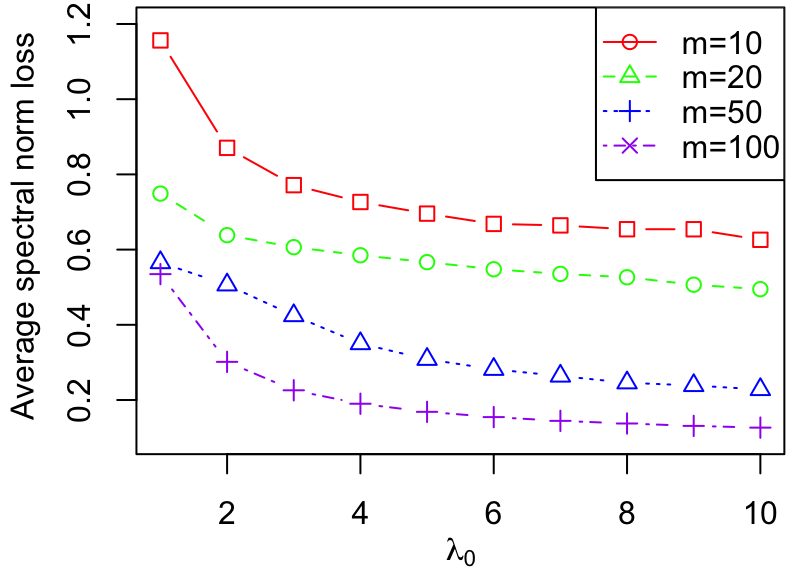}		
		\caption{Average Frobenius and spectral norm loss under the settings of Poisson observations. 
		}
		\label{fig:simu-setting4}
	\end{center}
\end{figure}

The results with 1000-times simulations under setting 1 - 2 are given in Figures \ref{fig:simu-setting3} - \ref{fig:simu-setting4}, repectively. 
From Figure~\ref{fig:simu-setting3}, we observe that the low-rank assumption of $A$ used in our method may not be necessary when $A$ is close enough to low-rank, i.e., when the singular values decay sufficiently fast as $\alpha$ increases. 
Besides, we find that as the intensity $\lambda_0$ increases in Figure~\ref{fig:simu-setting4}, i.e., the signal-to-noise ratio becomes stronger, we achieve more accurate matrix completion using \textsc{Macomss}.

Furthermore, we explore the missing-not-at-random mechanism for data generation. 

\begin{enumerate}
    \item[3.] Under Scenario~1 in Section~\ref{sec: DST}, we modify the block-missingness on the rows to be based on the row $L^2$ norm of the matrix. Specifically, after generating a matrix, we rearrange its rows in increasing order of their $L^2$ norms, and then remove the final 60\% of rows to construct the block-missingness. In this way, the probability of missingness depends directly on the unobserved values.
\end{enumerate}

\begin{figure}[h]
    \begin{center}
    \includegraphics[scale = 0.7]{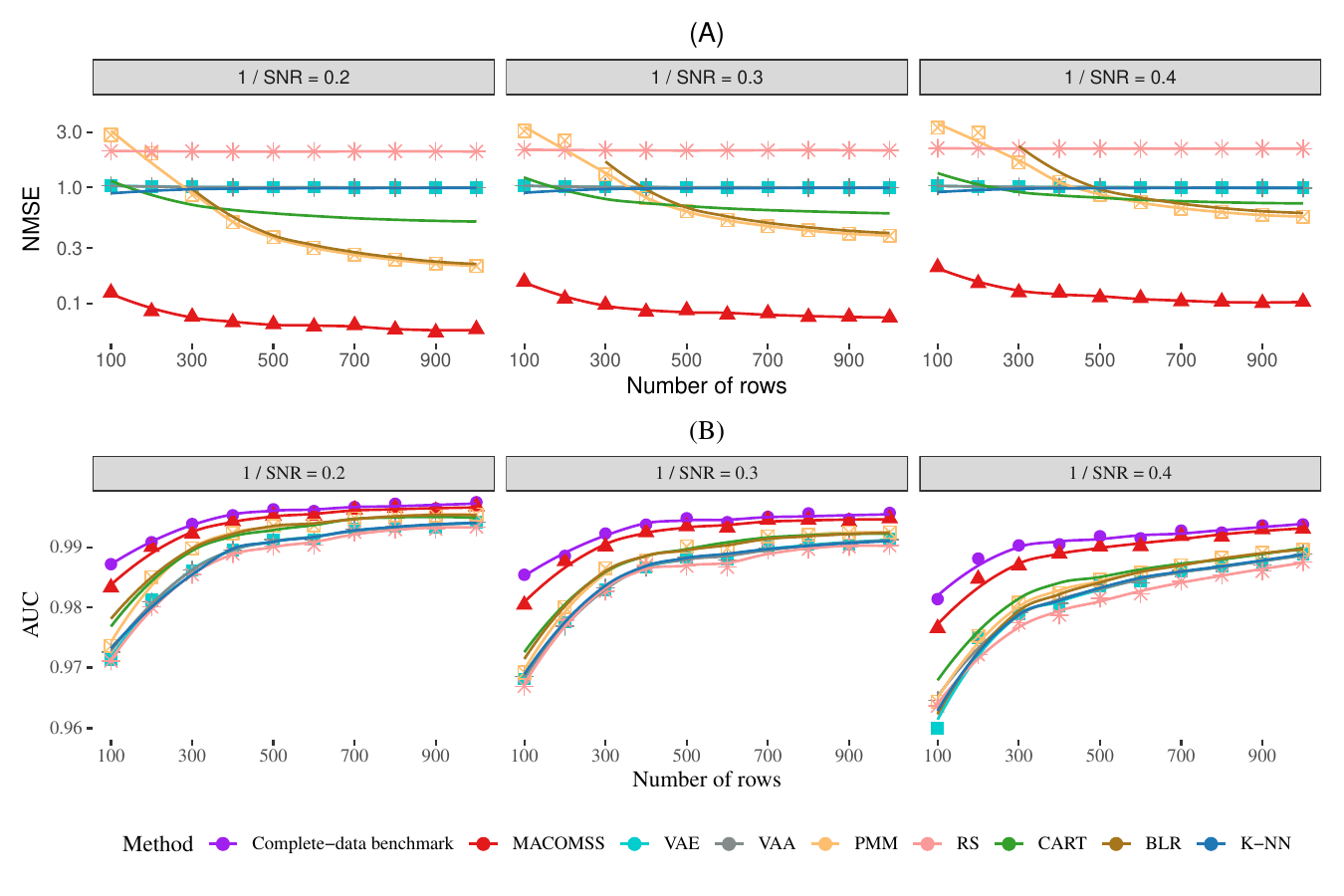}
    \caption{The $\text{NMSE}$ (A) and the $\text{AUC}$ (for simulated binary outcomes) (B) for scenario 1 for different numbers of rows from different methods.}
    \label{fig:performance-compare-1_SM}
    \end{center}
\end{figure}

Under this missingness mechanism, we compare \textsc{Macomss} with PMM, BLR, RS, CART, K-NN, VAE, and VAA, using the same settings as in Section~\ref{sec: DST}. 
The results are presented in Figure~\ref{fig:performance-compare-1_SM}. 
We observe that the performance of \textsc{Macomss} is similar to that shown in Figure~\ref{fig:performance-compare-1} and is superior to that of other methods. 
This demonstrates a certain robustness of \textsc{Macomss} under the MNAR mechanism.

\subsection{Generation Settings of Logistic Models}\label{sim_add}
For data generation, we first independently sample three matrices \( U \in \mathbb{R}^{n \times r} \), \( D \in \mathbb{R}^{r \times r} \), and \( V \in \mathbb{R}^{p \times r} \), where the entries of \( U \) and \( V \) are drawn independently from the Gaussian distribution \( \operatorname{Gau}(0, 1) \), and \( D \) is a diagonal matrix with its \( r \) diagonal elements drawn independently from the Gamma distribution \( \operatorname{Gamma}(2,2) \). 
For each $n$ and $p$, we set that \( r = 5 \). 
After that, we construct \( UDV^\top \) and standardize its columns to obtain the matrix \( X \).
Accordingly, we draw the \( i \)-th element of \( Z \), denoted \( Z_i \), from a Bernoulli distribution with the probability
\[
\mathbb{P}\{Z_i = 1\} = \frac{1}{1 + \exp(-\eta)},
\]
where \( \eta \) is associated with the predictor matrix \( X \) through \( \eta = \beta_0 + X_i^\top\beta_1 \), with \( X_i \in \mathbb{R}^p \) being the \( i \)-th row of \( X \), and \( \beta := (\beta_0, \beta_1^\top)^{\top} \in \mathbb{R}^{p+1} \) being the regression coefficient. The intercept \( \beta_0 \) is sampled from \( \operatorname{Gau}(0,1) \). Additionally, we randomly generate 15 entries of \( \beta_1 \) from \( \operatorname{Gau}(0,1) \), while the remaining entries are set to zero.

\subsection{Examination of the Low-rankness Assumption}\label{sec: low-rank}

\begin{figure}[h]
    \centering
    \includegraphics[width=0.6\linewidth]{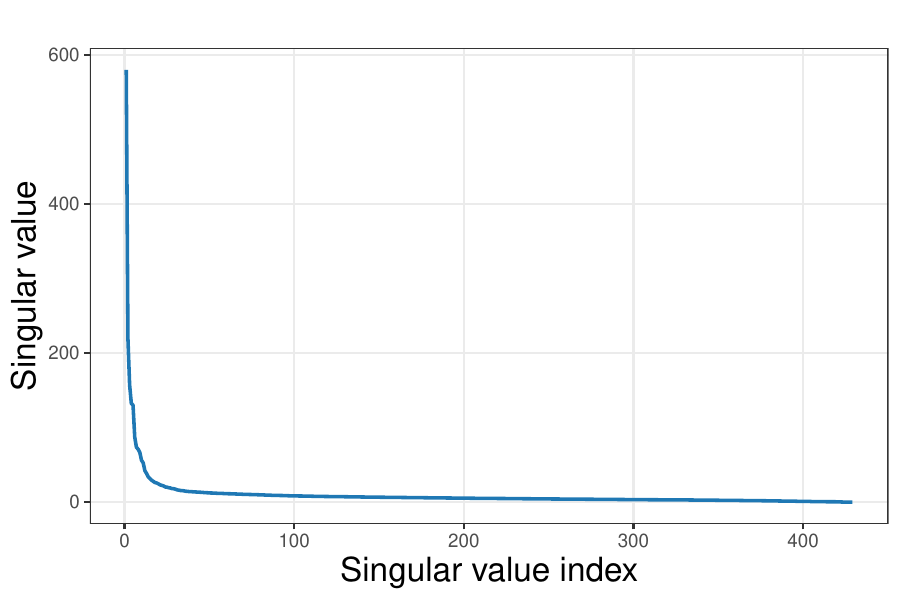}
    \caption{Singular value decay of the EHR data matrix with dimensions \(126{,}579 \times 438\).}
    \label{fig:singular_value_decay}
\end{figure}

To examine the low-rankness of the data matrix, we first normalize each variable in the EHR data matrix of dimension \(126{,}579 \times 438\) obtained from three hospitals. We then apply singular value decomposition to the resulting matrix and inspect the decay of its singular values. The results, shown in Figure~\ref{fig:singular_value_decay}, display a pronounced decay in singular values, indicating that the data matrix is approximately low-rank.

\end{sloppypar}

\end{document}